\def\etal{{\it et al.}}
\def\simlt{\hbox{ \rlap{\raise 0.425ex\hbox{$<$}}\lower 0.65ex\hbox{$\sim$} }}
\def\simgt{\hbox{ \rlap{\raise 0.425ex\hbox{$>$}}\lower 0.65ex\hbox{$\sim$} }}
\def\lta{\mathrel{\spose{\lower 3pt\hbox{$\sim$}}\raise 2.0pt\hbox{$<$}}}
\def\gta{\mathrel{\spose{\lower 3pt\hbox{$\sim$}}\raise 2.0pt\hbox{$>$}}}
\def\butabook{Buta, R., Crocker, D. A., \& Elmegreen, B. G. eds., 1996, 
``Barred Galaxies'', PASP, 91}
\def\msun{ \rm {M_\odot}}
\def\Msol{{M_\odot}}
\def\ten#1{\cdot 10^{#1}}

\documentstyle[aas2pp4,epsf,astrobib,rotate]{article}

\lefthead{Gyuk et al.}
\righthead{LMC Self Lensing}

\begin{document}

\title{Self-Lensing Models of the LMC}

\author{
G. Gyuk, N. Dalal, \& K. Griest
	}
\affil{Physics Department, University of California, San Diego, CA 92093}
\date{\today}
\centerline{July 23, 1999}
\begin{abstract} 
All of the proposed explanations for the microlensing events observed
towards the LMC have difficulties.  One of these proposed
explanations, LMC self-lensing, which invokes ordinary LMC stars as
the long sought-after lenses, has recently gained considerable
popularity as a possible solution to the microlensing conundrum.  In
this paper, we carefully examine the set of LMC self-lensing models.
In particular, we review the pertinent observations made of the LMC,
and show how these observations place limits on such self-lensing
models.  We find that, given current observational constraints, {\it
no} purely LMC disk models are capable of producing optical depths as
large as that reported in the MACHO collaboration 2-year analysis.  Besides
pure disk, we also consider alternate geometries, and present a
framework which encompasses the previous studies of LMC self-lensing.
We discuss which model parameters need to be pushed in order for such
models to succeed.
For example, like previous
workers, we find that an LMC halo geometry may be able to explain the
observed events.  However, since {\it all} known LMC tracer stellar
populations exhibit disk-like kinematics, such models will have
difficulty being reconciled with observations.  For SMC self-lensing, we
find predicted optical depths differing from previous results, but more than
sufficient to explain all observed SMC microlensing.  In contrast, for
the LMC we find a self-lensing optical depth contribution between 
$0.47\ten{-8}$ and $7.84\ten{-8}$, with $2.44\ten{-8}$ being the value 
for the set of LMC parameters most consistent with current observations.

\end{abstract}

\keywords{microlensing, dark matter, MACHOs, Magellanic Clouds,
galaxies: (halos, kinematics, dynamics)}

\section{Introduction}

Gravitational microlensing has become a powerful tool for the
discovery, limiting, and characterization of populations of dark (and
luminous) objects in the vicinity of the Milky Way.  Of great
interest is the interpretation of the handful of events discovered
towards the Magellanic Clouds.  If these events are due to a
population of objects in an extended Milky Way halo, they can be
interpreted to represent between 20\% and 100\% of the dark matter in
our Galaxy \cite{LMC2,gatesgyukturner}.  However, the most probable
masses of these objects lie in the 0.1 to $1 \msun$ mass range 
\cite{LMC2}. Such a large number of objects in this mass 
range is quite problematic (e.g. Fields, et al. 1998).  Therefore alternatives
to MW halo lensing have been sought to explain the LMC microlensing events.

One alternative, first proposed by Sahu (1994a), suggests
that stars within the LMC itself, lensing other LMC stars, could
produce the observed optical depth.  This claim has been disputed by
several other groups \cite{gould,LMC2}, who claim that the rate
of LMC self-lensing is far too low to account for the observed rate.
It was hoped that observation along a different line of sight
(i.e. towards the SMC) would resolve this issue.  After 5 years of
monitoring, there have been two observed microlensing events towards
the SMC.  The more recent SMC event was a resolved binary lens event
\cite{machosmc}, allowing determination of the lens distance
\cite{machosmc,erossmc,planetsmc}.  The lens was found to lie, with high
probability, in the SMC and not in the Milky Way halo.  There is also
evidence that the only other SMC microlensing event \cite{machosmc1} may
reside in the SMC \cite{erossmc1}. Thus all of the relevant lenses whose
distances are known are thought to reside in the Magellanic Clouds.  This
has been interpreted by some as settling the case in favor of the LMC/LMC
self lensing interpretation of the LMC events.  This conclusion is not
well-founded if based solely on the SMC events.  The reason, as we discuss
below, is fairly simple -- the SMC is known to be extended along the
line-of-sight, while there is little evidence that the LMC is similarly
extended.  In fact, the observations imply that the LMC is distributed as
a thin disk, quite unlike its smaller sibling.  Thus, unfortunately, the
SMC microlensing events do not settle the question of the interpretation
of the LMC events, and the controversy remains.

The purpose of this paper is to provide a set of calculations of LMC
microlensing that treats LMC self-lensing in a systematic, thorough
fashion.  We relate the known LMC observations to microlensing
predictions, and provide a framework in which future observations will
easily translate into microlensing predictions.  We hope this will serve
as a general basis for comparison between observation and theory in the
future.

Overall, we find that self lensing models typically suffer two major defects.
First, it is quite difficult for such models to produce enough
lensing to account for the observed optical depth, while remaining
within the bounds set by observation.  Second, the optical depth due
to disk or bar self-lensing is strongly concentrated on the sky, in
contrast to the rather uniform distribution of events seen to date.
These two statements have a major caveat: if the LMC lenses are
distributed in an extended or halo-like geometry, it is possible to
produce the required optical depth, and the central concentration of
the predicted events is significantly diminished.  Such an extended or
halo-like distribution, however, requires either an hitherto
undetected stellar population, or a dark MACHO component to the LMC
halo.  If a dark LMC halo is invoked, then one might expect it to have
a similar fraction of dark MACHOs as the Milky Way Halo.  Otherwise,
the presence of such a component in the LMC but not in the Galactic
halo would be puzzling.  On the other hand, if a stellar LMC halo with
a luminosity function similar to the disk is invoked, direct
observation of these LMC halo stars should be possible.  Indeed, as we
review below, several stellar populations which correspond to stars
that do trace the spheroid in our Galaxy have been observed in the
LMC, and {\it all} of them fail to exhibit a halo geometry.
Therefore, current observations suggest that the number of stars in
any such stellar halo is small, and that an LMC stellar halo probably
does not greatly contribute to microlensing. 

\section{Microlensing}

The first and main reason that previous work has produced such
discordant results is that different papers have treated the LMC
differently.  For example, Gould (1995) and Alcock et al. (1997) treated
the LMC as a thin exponential disk, while Sahu (1994a) and
Aubourg et al. (1999) modeled the LMC as being much more extended along the
line of sight.  These two qualitatively different prescriptions give
wildly different predictions for the optical depth and rate of
self-lensing.  The reason for this is simple.  The rate of
microlensing is proportional to the Einstein radius of the lenses,
which is given by
$$R_{\rm E} = \left[2R_{\rm S}{{D_{OL}D_{LS}}\over{D_{OS}}}\right]^{1/2}$$
where $R_{\rm S}$ is the Schwarzschild radius of the lens, $D_{OL}$ is the
angular diameter distance between the observer and lens, $D_{LS}$ is the
distance between the lens and source, and $D_{OS}$ is the distance between
observer and source.  The Einstein radius (and thus the microlensing rate)
tends to zero as the lens and source approach each other.  In the language
of Griest (1991), the ``Einstein tube'' pinches off at the ends.
Therefore, if the lenses are confined to a thin plane along with the
sources, the microlensing rate must be small.  On the other hand, if the
lenses are allowed to move away from the sources, the rate increases.
This principle is clearly demonstrated in the SMC.  Due to its
interactions with the LMC and the MW, the SMC is being tidally disrupted
and is consequently quite elongated along the line of sight to the MW
\cite{caldwell,welch}.  This allows stars within the SMC to be along the same
line of sight to us, but separated from each other.
Consequently, we expect appreciable self-lensing within the SMC, and this
expectation is borne out by the large observed SMC self-lensing rate
\cite{machosmc1,erossmc1}.  Indeed, EROS 2 reports an observed SMC optical
depth of $\sim 3.3\ten{-7}$ \cite{erossmc1}, and employing the simple model 
Palanque-Delabrouille et al. use to describe the SMC disk, we find
predicted self-lensing optical depths of 1.5, 3.0, and $4.4\ten{-7}$ for
SMC vertical scale heights of 2.5, 5.0, and 7.5 kpc respectively.  
(We note that these numbers do not agree with the predicted
optical depths that Palanque-Delabrouille et al. report, but we are
confident that our values are correct.)

To answer the question of whether LMC self-lensing is significant, we must
understand the distribution of stars within the LMC.  If the LMC is a thin
disk, then the small rates and optical depths derived by Gould and others
will be valid.  Conversely, if the LMC is puffy, then the large rates and
optical depths claimed by Sahu and others will be correct.  The basis for
any description of the LMC is the set of observations that have been made
of the LMC. We therefore turn to the current state of observations of the
LMC.

\section{Observations and Models of the LMC}

\subsection{LMC disk}
Since the pioneering work of de Vaucoleurs (1957), it has been well
accepted that the stellar component of the LMC has an exponential
profile.  The value de Vaucoleurs measured for the exponential scale
length, $R_d$, continues to agree with the current value of
$1.8^\circ$ \cite{MACHO9million}, 
which corresponds to a physical scale length of 1.6 kpc
for a distance to the LMC of 50 kpc.  In addition
to this stellar population, the LMC possesses significant quantities
of HI gas, which has recently been mapped out by Kim et al. (1998).
Their images show clear spiral structure in the gas, supporting the
notion that the LMC is a typical dwarf spiral galaxy.  The gas is
confined to a thin disk, inclined at roughly $30^\circ$, with a
position angle $\sim 170^\circ$.  See Westerlund (1998, p. 30) for a 
compilation of various estimates of the LMC orientation, as well as 
Kim et al. (1998) for a recent value.  Based on these observations, in this
paper we describe the stellar disk by a double exponential profile,
given by
$$\rho_d={M_{disk}\over{4\pi z_d R_d^2}} e^{-{R\over{R_d}}-|{z\over{z_d}}|},$$
where $R_d$ is the radial scale length, $z_d$ is the vertical scale
height, and $M_{disk}$ is the disk mass.  Note that $R_d$ is well constrained
by observation, but we have some leeway in the scale height and in the
mass of the disk. We discuss these two quantities in more detail later. The
disk is inclined at angle $i$ to our line of sight and has position angle
PA.

\subsection{LMC Bar}

As is well known, the LMC hosts a prominent bar, of size roughly
$3^\circ\times 1^\circ$.  The bar has the unusual (although not
unique, e.g. Freeman 1996, Odewahn 1996) property of being offset from
the dynamical center of the HI gas.  The offset is $\approx 1.2^\circ$
\cite{westerlund}, corresponding to a physical offset of $\sim 1$ kpc.
The kinematics of the LMC bar are consistent with solid body rotation
\cite{odewahn96}, as is seen in numerous barred galaxies.  The
distribution of matter within the bar is not well known.  Measurements
of the luminosity function, after subtraction of disk light, show it
to be consistent with an exponential profile along the major axis
\cite{bothun88,odewahn96}.  This is consistent with certain other
bars, which can be well described by an exponential along the major
axis and a Gaussian profile along the minor axis
\cite{blackman83,ohta96}.  For our own Galactic bar,  Dwek, et al. (1995) 
have proposed a profile similar to a Gaussian, but more boxy, and this 
form is also consistent with bars in certain other galaxies.

Thus, unlike the disk, the bar is not particularly well defined.  With little
guidance from observations, we have treated the bar simply as a triaxial
gaussian, with axis ratios chosen to match the observed ratios.  We let
$$\rho_b = {M_{bar}\over{(2\pi)^{3/2} x_b y_b z_b}} e^{-{1\over 2}
[({x\over x_b})^2+({y\over y_b})^2+({z\over z_b})^2]},$$
where $x,y,z$ are coordinates along the principal axes of the bar, and
$x_b,y_b,z_b$ are the scale lengths along the three axes.  $M_{bar}$ is the 
total mass of the bar.  This form is somewhat
similar to models used to describe other galactic bars, e.g. \cite{dwek}.
We place the bar in the same plane as the disk, however we place the bar
center at the position of the observed bar centroid, at ($\alpha=5h24m,
\delta=-69^\circ 48'$) \cite{deVauc}.  We use a position angle for the bar 
of $120^\circ$.

It is not clear, at this point, how great an influence the bar exerts over
the dynamics of the surrounding gas and stars.  In many barred galaxies,
the bars sweep up gas and drive it towards the center
\cite{kenney96,ho96,nick99}.  In our own Galaxy, the kinematics of gas in the
inner regions is strongly influenced by the putative bar \cite{weiner99}.
While there is evidence for non-axisymmetric
flows in the vicinity of the bar \cite{dottori96,odewahn96}, indicating
that the bar may be dynamically important, the HI maps of Kim et al. show that 
the LMC bar does not dominate
the central dynamics.  From this we conclude that the mass of the bar 
cannot exceed the disk mass in the central regions, which leads us to a 
bound on the bar mass.  Now, $\sim 25\%$ of the disk mass lies within one 
scale length, while most of the bar lies in this same central region.
We thus arrive at the following restriction: $M_b < 25\% M_d$ to avoid
bar domination.  This agrees nicely with the estimates of Sahu (1994b), 
who suggested a bar to disk mass ratio in the range 15-20\% 
based on luminosity considerations. 

\begin{figure*}
\epsfysize=20.0cm
\centerline{
\rotate[r]{\epsfbox{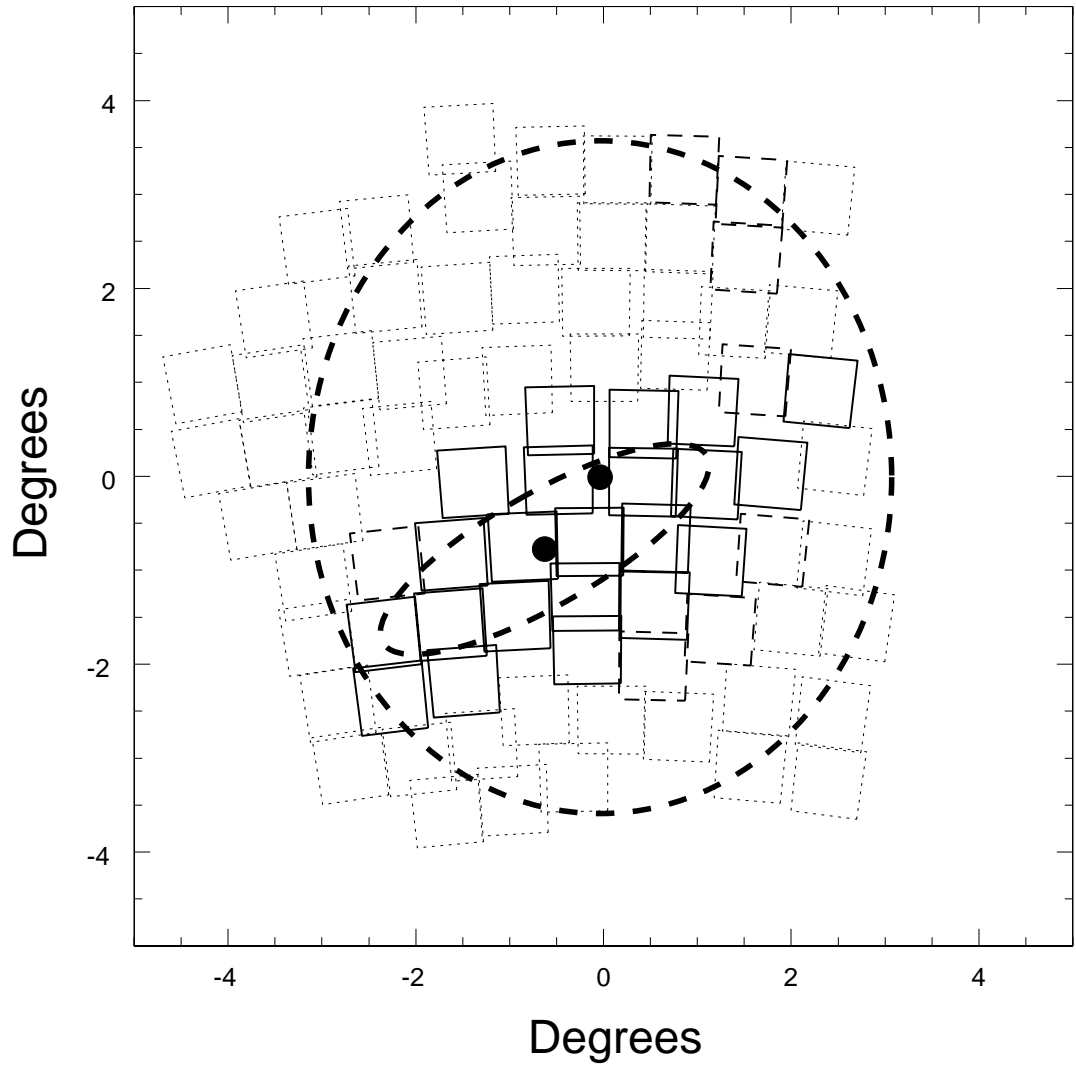}}}
\caption{
MACHO fields.  The solid outline squares depict the 22 MACHO fields reported
in Alcock et al. (1997a).  The dashed squares are the 8 additional
fields MACHO will report in their year 5 paper, and the dotted squares
are the other 52 fields that they monitor.  The thick dashed ellipses
show the position and orientation of our model disk and bar.  They are
plotted at two scale lengths.
\label{figfields}
}
\end{figure*}

\subsection{LMC Velocity Distribution and Vertical Scale Height}
Now we turn to the velocity distribution of the model stars. Perhaps
the best determination of the inner velocity curve of the LMC is in
the work of Kim et al. (1998).  When supplemented by outer rotation
curves derived from carbon stars \cite{kunkel} and clusters and
planetary nebulae (PNe) \cite{schommer} the basic outline is clear:
the circular velocity rises rapidly in the first two kpc and then
levels off and is flat at about 70 km/s out to at least 8 kpc. There
are indications of a possible dip at 3 kpc though this may not be
significant. For our models we approximate the rotation curve by solid
body rotation out to a radius $r_{\rm solid}=2$~kpc, followed by flat
rotation at $v_c=70$~km/s.

\begin{table*}
\begin{tabular} {llcl}
Population & Study & Velocity Dispersion &Age \\
\hline
supergiants	& Prevot, Rousseau \& Martin (1989)&  6   &young\\
HII	   	& ~~~~~~~~~~~~~``                  &  6   &young \\
HI         	& Hughes et al. (1991)             &  5.4 &young\\
VRC        	& Zaritsky \& Lin (1997)           & 18.4 &young?\\
PNe        	& Meatheringham et al. (1988)      & 19.1 &intermediate\\
OLPV       	& Hughes et al. (1991)             & 33   &old\\
ILPV		&	~~~~~~~~~~~~~``		   & 25   &intermediate\\
YLPV		&	~~~~~~~~~~~~~``		   & 12-15&young\\
OLPV       	& Bessel et al. (1986)             & 30   &old\\
metalpoor giants& Olszewski et al. (1993)          & 23-29&old\\
metalrich giants& ~~~~~~~~~~~~~``                  & 16.0 &intermediate?\\
new clusters	& Schommer et al. (1992)	   & 20   &intermediate\\
old clusters  	& ~~~~~~~~~~~~~``		   & 30   &old\\
carbon stars	& Kunkel et al. (1997)		   & 15   &young\\
CH stars (disk)	& Cowley \& Hartwick (1991)	   & 10   &yng/intermed?\\
CH stars (halo)	&	~~~~~~~~~~~~~``		   & 20-25&old\\ \hline
\end{tabular}
\caption{Observed velocity dispersions for various populations.}  
\end{table*}

These studies help to define to bulk motions of gas and stars in the LMC.
However, as Gould (1995) has shown, velocity dispersions (and the implied
scale heights) are crucial in determining the optical depth and rate of
microlensing.  So let us consider measurements of the velocity dispersion
of LMC populations.  Prevot, Rousseau \& Martin (1989) studied late-type
supergiants and HII regions, concluding that the internal velocity
dispersion of this population is approximately 6 km/s. This is quite close
to that of the HI gas, 5.4 km/s (Hughes et al. 1991) which is hardly
surprising as these tracers all belong to a very young population. A somewhat
older population is probably illustrated by the disk-like ($\sigma_v\sim
10$ km/s) CH stars found by Cowley \& Hartwick (1991) which seem to
correspond to ``CH-like'' stars found in the Galactic disk
\cite{yamashita}.  Cowley \& Hartwick also found, however, a population
with a considerably higher velocity dispersion (20-25 km/s), that
presumably corresponds to an even older population.  Meatheringham et
al. (1988), in a study of planetary nebulae (PNe), found that the
intermediate population of stars represented by the PNe are rotating as
fast and in the same disk as the gas, but with a velocity dispersion of
19.1 km/s. Bessel et~al. (1986) observed old (age $\sim 10^{10}$ yrs) long
period variables (OLPV's), which are thought to trace the oldest stellar
populations, and obtained a mean line-of-sight velocity dispersion of
about 30 km/s.  Hughes et al. (1991) also observed OLPV's and obtained
similar results ($\sigma\sim 30$ km/s).  We note here, for future
reference, that a spheroidal distribution would require velocity
dispersions of $\sigma\approx v_c/\sqrt{2}\approx 50$ km/s.  Since the
observed dispersions of the OLPV's fall far short of this, we see that
even this oldest population derives much of its support from rotation,
and therefore exhibits ``disk-like'' kinematics.

The general trend among the many kinematic studies of the LMC seems to
be clear: tracers have velocity dispersions ranging from $\sim 5$ km/s
for very young ages to $\sim 30$ km/s for the most ancient
populations.  {\em All} LMC populations studied to date have disk-like
kinematics regardless of age \cite{olszewski}.  Table~1 lists some of
the more recent kinematic studies of the LMC by population type,
velocity dispersion and probable age.

From the velocity dispersions, let us now turn to the vertical scale heights.
Bessel et al. (1986)
estimated the vertical scale height of the oldest population
to be roughly 0.3 kpc, while Hughes et al. (1991)
estimated the scale height to
be $\simlt 0.8$ kpc.  They emphasized that this was the oldest population,
accounting for at most 2\% of the mass of the LMC. The majority of the LMC
disk, they contend, should possess a more compact vertical distribution
and smaller vertical velocity dispersions. This is supported by RR Lyrae
and cluster studies which suggest that the ancient extended populations
make up considerably less than 10\% of the LMC stars
\cite{olszewski,kinman}.  We thus allow our scale height (which should
characterize the bulk of the LMC population) to range up to 0.5 kpc
and adopt velocity dispersions in a corresponding range of 10-30
km/s. In theory these parameters should be tied together by the
vertical Jeans equation (however see Weinberg 1999).  In practice,
our knowledge of the total mass and mass distribution of the LMC is
poor enough that we simply note that the opposite extremes of these
ranges (i.e. 10 km/s with 0.5 kpc and 30 km/s with $< 0.2$ kpc) are
likely inconsistent.

\subsection{LMC Halos: Light and Dark}
The above distributions describe the known stellar populations.  We again
reiterate that the non-detection of any stellar halo population
($\sigma_v \sim 50$ km/s) places
severe constraints upon the existence of such a stellar halo.  The above
noted RR Lyrae and cluster studies, along with the OLPV observations limit
the stellar halo to perhaps 5\% of the mass of the LMC 
\cite{HWR,olszewski93,olszewski,kinman}. 

This, however, places no limits upon the existence of a dark halo, to
which we now turn.  Obviously, even less is known about the LMC dark
matter than its luminous populations.  As Kim et al. (1998)
discuss, the observed
LMC rotation curve is inconsistent with the distribution of known
populations, given the assumption of a {\it constant} mass to light ratio.
Schommer et al. (1992) obtain similar results.  Although the variation in
the mass to light ratio required to explain the rotation curve with
luminous matter alone is less than a factor of two, we can take these
results as prima facie evidence for dark matter within the LMC.  This
should not be too surprising, since studies of the velocity curves of
similar dwarf galaxies show that they are dominated by dark matter (see
e.g. Carignan \& Purton 1998).  Models without dark halo also exist that
can explain the rotation curves (D. Alves 1999, private communication),
but these will not be discussed here.  There are numerous models that have
been used to describe dark matter in galaxies.  Most common, and perhaps
easiest, is the simple spherical pseudo-isothermal distribution,
$$\rho_h=\rho_{0} \left[1+{{r^2}\over{a_h^2}}\right]^{-1}$$
with core radius $a_h$ and central density $\rho_{0}$.  In the limit
$a_h\rightarrow 0$, this distribution gives an isothermal Maxwellian
velocity profile \cite{BT}.  For the LMC, however, core radii smaller than
$a_h \sim 1$\ kpc lead to problems matching the rotation curve.  
Although not self-consistent, for simplicity we use $a_h>1$ kpc and a 
uniform Maxwellian velocity distribution.  We take the
fraction of this halo in MACHOs to be $f_M$. 

Since the LMC is embedded in the (dominating) gravitational potential
of our Galaxy, we expect the LMC to have a tidal radius, beyond which
objects are not stably bound to the LMC \cite{BT}.  This places a
limit on the size of the LMC halo.  Although the density should
smoothly decline to zero near the tidal radius, for simplicity we
instead implement a truncation radius, $r_t$, beyond which the LMC
halo density abruptly vanishes.  Using star counts from the 2MASS
survey, Weinberg (1998) has estimated $r_t\approx 11$ kpc.  This is
the value we adopt.

\begin{table*}
\begin{tabular} {lrlll}
Study & Mass Estimate & Radius & Component&\\
\hline
Hughes et al. (1991)	& $6.0 \cdot 10^9 \Msol$ 	& 4.5 kpc & Total&
Spheroidal estimator\\
Kim et al. (1998)	& $2.5 \cdot 10^9 \Msol$	&         & Disk&
Rotation curve fit\\
~~~~~``		& $3.4 \cdot 10^9 \Msol$	& 8 kpc	  & Halo&\\
Schommer et al. (1992)	& $\sim2.0 \cdot 10^{10} \Msol$    & 5 kpc	  & Total&
Spheroidal estimator\\
~~~~~``		& $1.0 \cdot 10^{10} \Msol$	& 8 kpc   & Total&
Rotation estimate\\
Meatheringham et al. (1988)& $3.2 \cdot 10^9 \Msol$	&	  & Disk&
Rotation maximum fit \\
~~~~~``		& $6.0 \cdot 10^9 \Msol$	& 5 kpc	  & Total&\\
Kunkel et al. (1997)	& $6.2 \cdot 10^9 \Msol$	& 5 kpc	  & Total&
Point mass estimation \\
~~~~~``                & $<1.0 \cdot 10^{10} \Msol $	&         & Total&\\
 \hline
\end{tabular}
\caption{Estimates of the LMC mass.  Note that some entries refer to 
specific components, such as the disk or halo, while other entries 
correspond to the LMC as a whole.}
\end{table*}
\subsection{LMC Mass}
\begin{figure}
\epsfysize=4.85cm
\centerline{
\epsfbox{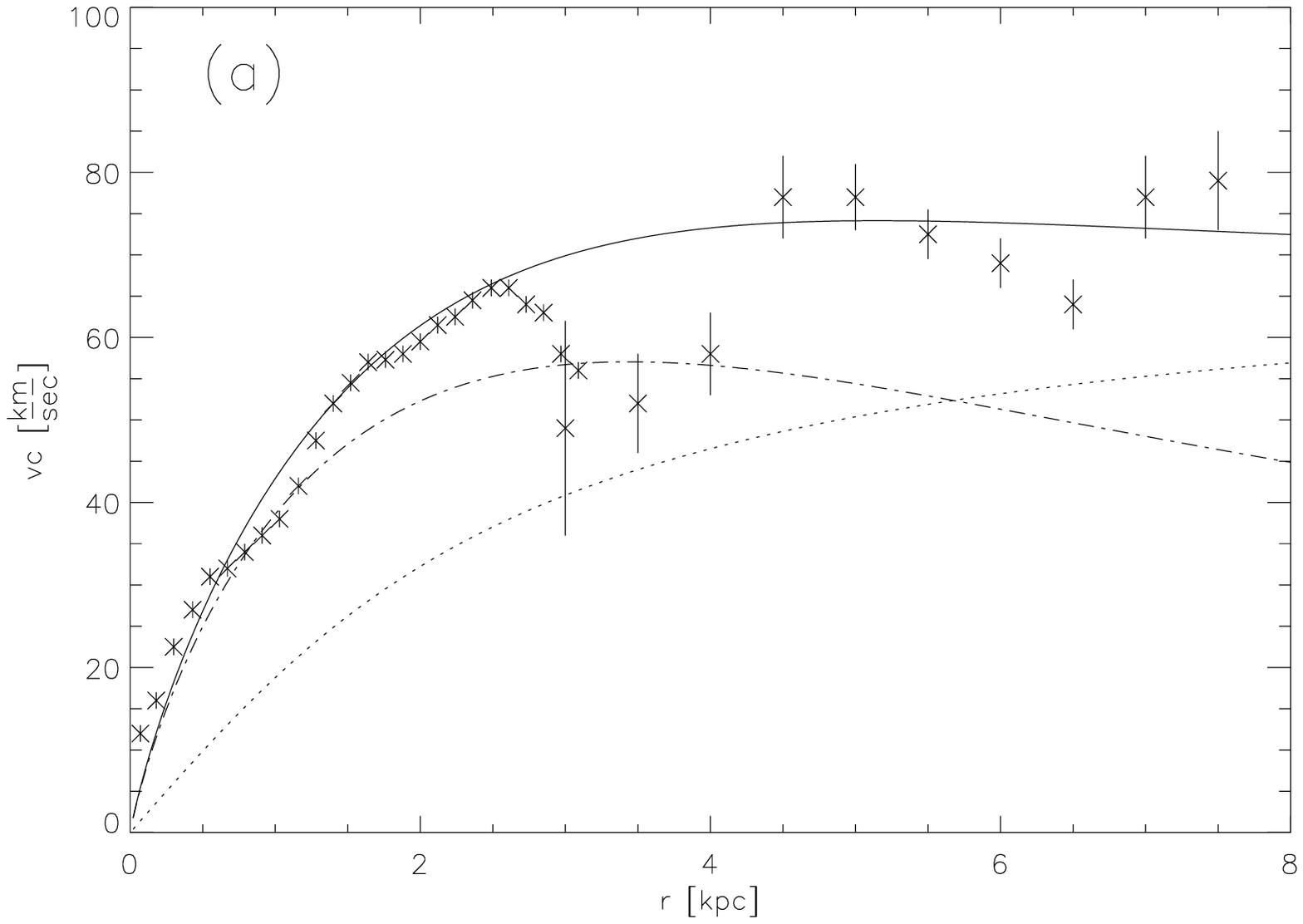}}
\epsfysize=4.85cm
\centerline{
\epsfbox{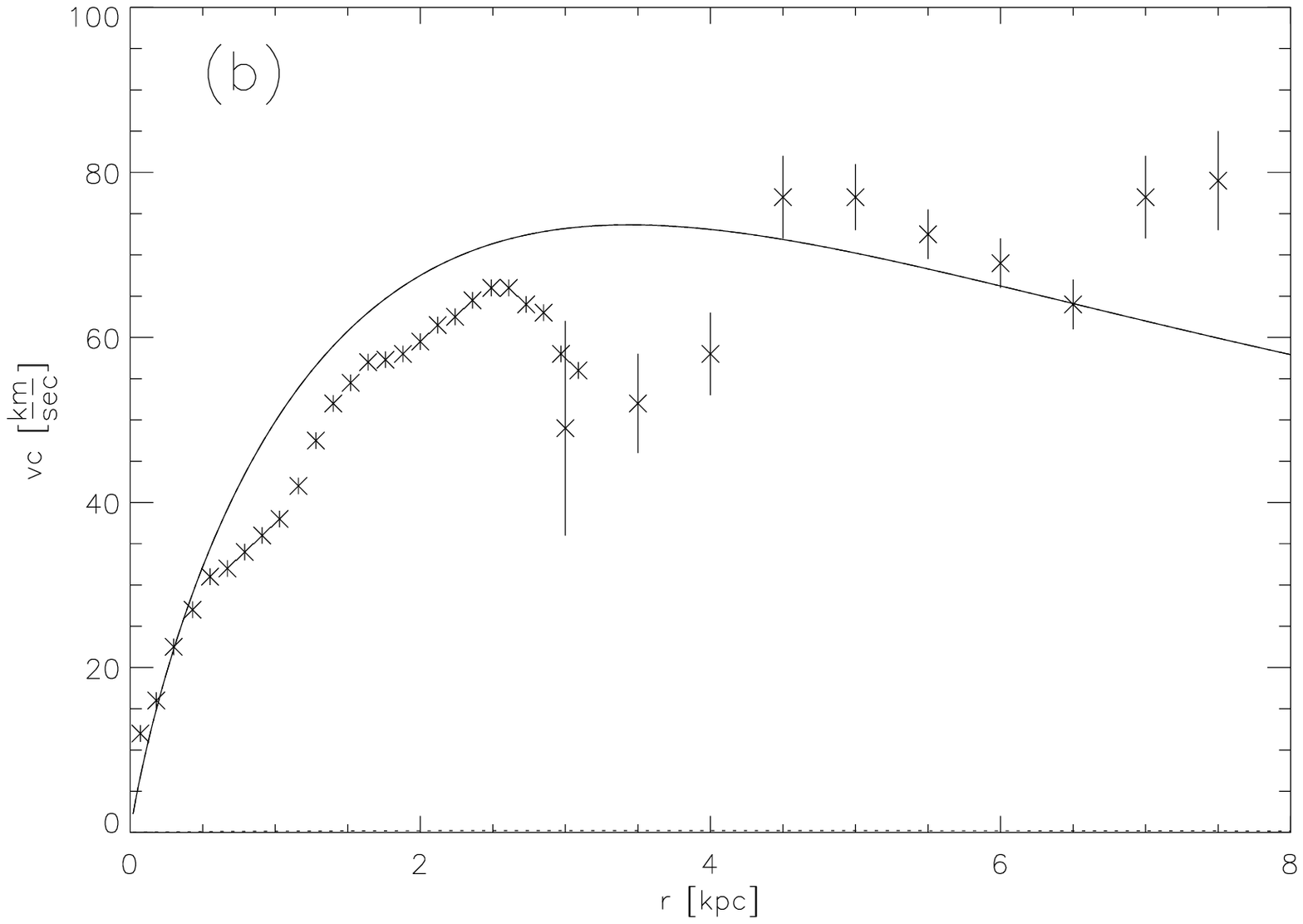}}
\epsfysize=4.85cm
\centerline{
\epsfbox{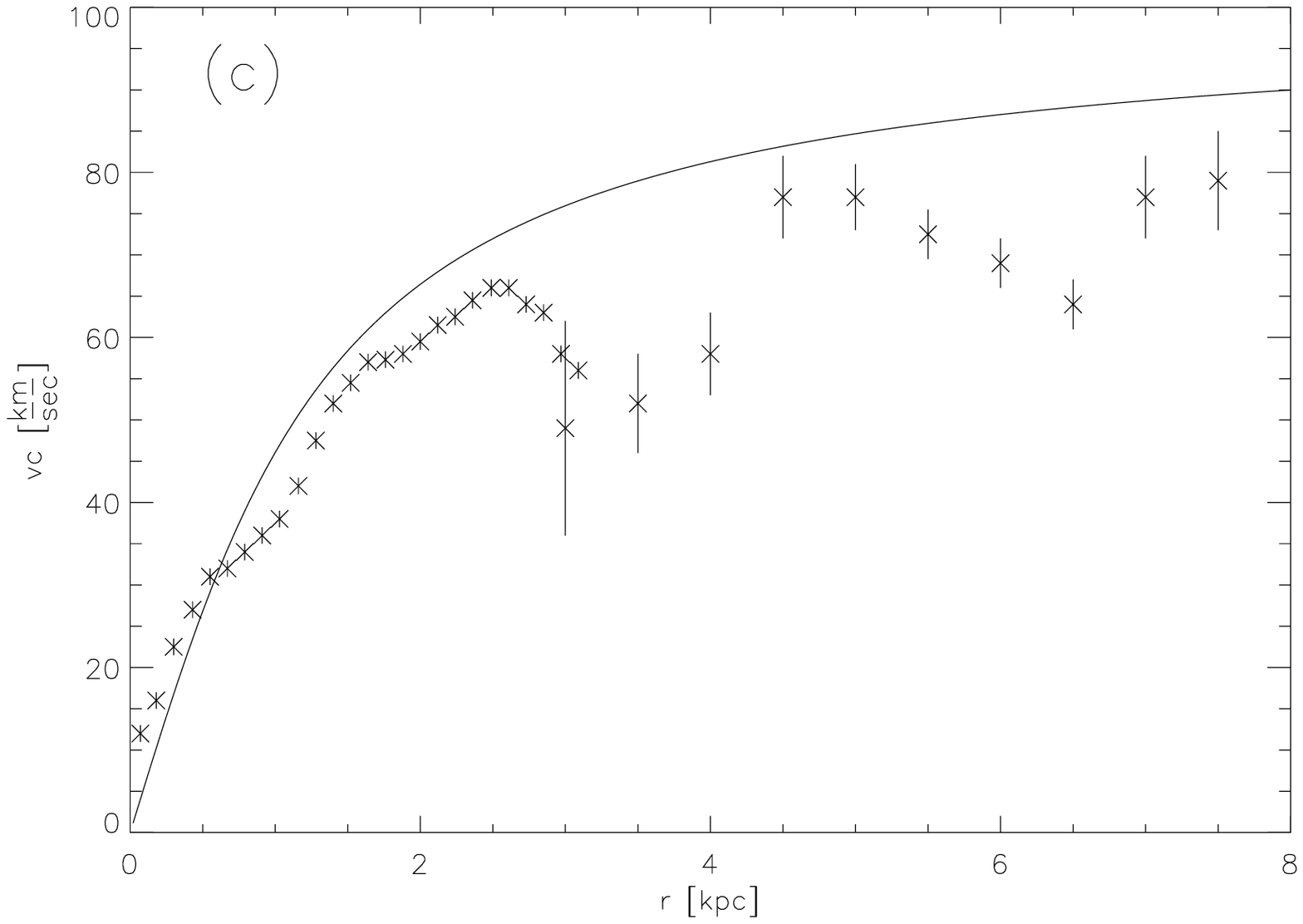}}
\caption{
Model rotation curves.  The plotted points with error bars are from 
Kim et al. (1998) and Kunkel et al. (1997).
Panel (a) shows the predicted rotation curve for
our preferred model, with $M_{disk}=3\ten{9}M_\odot$ (entirely in the
double exponential disk), $M_{dark}$(8 kpc)=$6\ten{9}M_\odot$ and 
$a_h=2.0$\ kpc.  The dash-dot line shows the disk contribution, and the
dotted line shows the halo contribution.  Panel (b) shows the predicted
rotation curve for the maximal disk we allow, $M_{disk}=5\ten{9}M_\odot$.
Note that it already significantly overshoots the observed velocities.
Any additional dark component exacerbates the problem.
Panel (c) shows the curve for the maximum dark halo mass, 
$M_{dark}$(8 kpc)=$1.5\ten{10}M_\odot$ and $a_h=1$\ kpc.  Clearly,
masses in excess of this can safely be ruled out.
\label{figrotcurve}
}
\end{figure}

The question of the LMC mass is an unsettled one. A few of the more
recent mass estimations are shown in Table~2.  Estimates range from
only a few $\times 10^9 \Msol$ to $\sim 2 \cdot 10^{10}\Msol$.  Close
inspection, however, reveals a few regularities. The highest estimates
are based of the spheroidal estimator of Bahcall \& Tremaine (1981),
which assumes both velocity isotropy and a spherical mass
distribution.  Since both of these conditions are likely to be
violated, the spheroidal estimators should be taken as upper limits. A
similar argument can be made for the point mass estimation of Kunkel
et al. (1997). With these caveats in mind the data seem consistent
with a disk of perhaps $3 \cdot 10^9 \Msol$ and a halo whose mass
within 8 kpc is roughly $6 \cdot 10^9 \Msol$.  While the extremely
high quality HI data of Kim et al. (1998) would appear to rule out a
disk mass much in excess to this, the halo component is much more
uncertain. We thus take $M_{disk}+M_{bar} \le 5.0 \cdot 10^9 \Msol$ and allow
the total LMC halo mass within 8 kpc to range up to $1.5 \cdot 10^{10}
\Msol$.  In figure \ref{figrotcurve} we show the rotation curves associated
with several choices of component masses.  While our preferred parameters
fit the observed velocity profile quite nicely, the upper ends of our allowed
ranges are clearly starting to run afoul of the observations.

This covers all of the populations we consider for LMC self lensing.
The various parameters and their preferred values 
are listed in Table~3. However, since observations are subject to change,
it is worthwhile to consider not only the currently preferred description
of the LMC, but a wide class of models and parameters, so that future
observations easily translate into microlensing predictions.
We have therefore also indicated the acceptable range of each model
parameter in Table~3.  
Of course, models that simultaneously take extreme values of
all the parameters may not be realistic.  While this range {\em spans} the set
of acceptable models, not all models in this range are acceptable.

\begin{table}
\begin{tabular} {lcl}
\hline
Parameter & Preferred Value& Allowed Range \\
\hline
inclination	& $30^\circ$	& $20-45^\circ$\\
$R_d$		& $1.8^\circ$	& $1.8^\circ$\\
$z_d$		& 0.3 kpc	& 0.1-0.5 kpc \\
$v_c$		& 70 km/s	& 60-80 km/s  \\
{\it L}		& 50 kpc	& 45-55 kpc   \\
$\sigma_v$	& 20 km/s	& 10-30 km/s  \\
$a_h$		& 2 kpc 	& 1-5 kpc     \\
$M_{d+b}$(8kpc)	& $3\ten{9}M_\odot$	& $<5\ten{9} M_\odot$ \\
$M_{dark}$(8kpc)	& $6\ten{9}M_\odot$	& $<1.5\ten{10}M_\odot$  \\
$M_{bar}/M_{d+b}$	& 0.15	& 0.05-0.25    \\
$M_{stellar~halo}$ & $0\ten{8}\msun$	& $0-5.0\ten{8}\msun$    \\
\hline
\end{tabular}
\caption{Model parameters.  $M_{d+b}=M_{disk}+M_{bar}$.  All masses
are for LMC distances of 50 kpc}
\end{table}

\section{Calculations}

Using the models specified in the previous section, the microlensing event
rate, optical depth, and timescale distribution can be calculated.  The
microlensing rate is the number of events per year per star. To obtain the
total number of events expected for an experiment one would multiply the
rate by the observational exposure, which is defined as the number of
monitored stars times length of time they were monitored.  The optical
depth is the probability that a given source star is lensed with
magnification greater than 1.34.

The optical depth along a given line of sight is given by 
$$\tau={1\over{N_s}}\int_0^\infty dL n_s(L)\int_0^{L}dl\pi 
R_{\rm E}^2(L,l) n_l(l)$$
where $n_s$ is the number density of sources, $n_l$ is the number density
of lenses, $R_{\rm E}$ is the Einstein radius defined above, $L=D_{OS}$ is
the distance to the source, and $N_s=\int n_s dL$ is the total number of
sources along the line of sight.  Writing $\rho({\vec x})=\langle m\rangle
n_l({\vec x})$ and inserting the expression for the Einstein radius
$R_{\rm E}$ gives
$$\tau={{4\pi G}\over{c^2N_s}}\int_0^\infty dL n_s(L)
\int_0^{L}dl{{l(L-l)}\over L}\rho_l(l).$$

Before proceeding further, we discuss the source distribution in more
detail. In the simplest approach the source density would be set to the
mass density (disk+bar) we have already discussed. This ignores two
important issues. First, the LMC is seen almost face-on, and hence the
thin dust and gas disk obscures the far half of the stars. This
preferentially removes source stars with the highest optical depth,
lowering the observed optical depth. Modeling the extinction as a zero
thickness plane of 0.4 V-magnitudes \cite{LMCred}, a rough approximation
shows that the effect should reduce the optical depth by about 15\% for
disk-disk self lensing. In the following we ignore this effect, simply
noting that our quoted results are over-estimates. The second issue
concerns the differing populations of the disk and bar. These different
populations (due to the varying ages and star formation history) yield
different source to mass ratios. Unfortunately, with the present knowledge
of the bar and disk luminosity functions and relative metallicities
etc. a precise calculation is impossible. We therefore assume a uniform mass 
to source ratio.
  
We have found it convenient to calculate the optical depth, event rate,
and duration distribution using a Monte Carlo method.  One advantage of
the Monte Carlo technique is that it easily allows consideration of
arbitrary spatial distributions of lenses and sources. Another important
advantage of the Monte Carlo method is the ease with which we were able to
average over the experimental fields as discussed in the following
section. In addition, the separate integrals for the optical depth, rate,
and event timescales can all be evaluated simultaneously, in one fell
swoop.

\subsection{MACHO Fields}
For self lensing models the optical depth varies rapidly with position
in the LMC. Thus a single number, ``the optical depth to the LMC'' is
only useful if the precise location of the observed sources is
specified. In order to make the comparison to the observed optical
depth we have chosen to average our results over the MACHO
collaboration fields \cite{LMC2}.  Ideally we should fold in the
experimental efficiency and relative source numbers in each field as
observed.  Since these numbers are unavailable, however, we have
weighted each field by the {\em model} number of stars.  We show in
Figure \ref{figfields} the fields over which the optical depth is
averaged.  The solid outlines depict the 22 fields covering about 10
square degrees used in the MACHO year 2 analysis \cite{LMC2}.  Note
that the year 2 fields are concentrated along the regions of highest
numbers of source stars.  The dotted outlines describe the roughly 40
square degrees (82 fields) that are being monitored by the MACHO
collaboration.\footnote {The centers of all 82 fields can be accessed
from the MACHO collaboration web site, at
http://wwwmacho.mcmaster.ca/} These cover most of the LMC disk out to
a radius of 3.5 degrees from the center (about 2 disk scale lengths).
Later we will discuss possible observational consequences of the
increased coverage.  Also of interest are the 30 fields that will be
presented in the MACHO collaboration 5 year analysis
\cite{vandehei98}.  We note that the EROS II collaboration is 
similarly observing most of the LMC disk, while the OGLE II 
collaboration monitors 4.2 square degrees \cite{ogle2}.

\subsection{Mass function}
Although the optical depth is independent of the mass function, the
event rate and the timescale distribution do depend upon the lens
masses.  We therefore must consider an appropriate mass function for
the lensing population.  As we expect the rate to be dominated by low
mass stars, we choose to employ the MF derived by Gould, Bahcall \&
Flynn (1997), which they based upon counts of M dwarfs in the MW disk.
Gould et al. found that
$${{dN}\over{dm}}\propto\left({M\over{0.59M_\odot}}\right)^\alpha,$$
with $\alpha\approx -0.56$ for $m<0.59M_\odot$ and $\alpha\approx
-2.21$ for $m>0.59M_\odot$. Since the timescales and rate are dependent
only on the square root of the masses the precise details of the LMC mass
function are not important unless it is radically different from that of
the MW. 

\section{Results}
The measured total optical depth towards the LMC 
from the 2-year MACHO collaboration analysis 
is $2.9^{+1.4}_{-0.9} \ten{-7}$ \cite{LMC2}.  
How does this compare with the predicted range of optical depths of the
above models?

\subsection{Disk/Bar Optical Depth and Scalings}

Let us first consider disk-disk lensing.  Evaluating the integral, and
plugging in the preferred disk parameters listed in Table~3, we obtain a
(22) field averaged optical depth of $\tau=1.46 \cdot 10^{-8}$ for
disk-disk lensing.  This is the best value for the disk-disk optical
depth, given the current status of observations.  As noted, however, we
should explore the dependence of $\tau$ upon the model parameters.
We can obtain a reasonable scaling using a simple-minded argument.
Let's first consider the optical depth for a single source star at the LMC 
center.  This integral is easy to do exactly, but for our purposes we are 
interested in its asymptotic behavior.  Writing the integral in dimensionless
form, we see it is approximately $\propto\rho_0\int e^{-Ax}xdx\propto\rho_0
A^{-2}$, where $\rho_0=M_{disk}/(4\pi R_d^2z_d)$ is the central density and
$$A=L\left({{\cos i}\over{z_d}}+{{\sin i}\over{R_d}}\right)$$
is the only other form in which these parameters enter the integral.
The extended source distribution will modify this scaling, but for
disk-disk lensing, the source distribution enters the line-of-sight integral
again in the form of $A$.  Thus, $\tau\sim\rho_0L^2F(A)$ should capture the 
essential behavior of the 3-d averaged optical depth, with asymptotic 
leading behavior $\tau\propto\rho_0A^{-2}$.  For the LMC, $A\approx 160$, 
so we expect the expansion
$$\tau\propto{{M_{disk}L^2}\over{R_d^2z_dA^2}}(1+a_1A^{-1}+a_2A^{-2}+...)$$ 
to describe accurately the scaling of the optical depth with the parameters
over the range of interest, even if we keep only the first one or two terms. 
However, for estimation purposes, the leading behavior will be good enough.

We arrive at an interesting relation if we further approximate this
already zeroth-order treatment.  For a nearly face-on, thin disk,
$A\approx L\cos(i)/z_d$.  Then $\tau\propto Mz_d/(R_d^2\cos^2i)$.
Note that a similar result was derived by Sahu \& Sahu (1998).  Now,
quantities such as $M, R_d$, etc. are derived, not measured directly.
They are inferred from measured parameters such as the apparent axis
ratio $k=\cos i$, the rotation curve (which itself is derived from
radial velocity measurements), the distance $L$, and the vertical
velocity dispersion $\sigma_z$.  For example, the vertical scale 
height $z_d$ is typically computed using the Jeans equations, which 
for a self-gravitating thin disk in equilibrium demand that 
$z_d\sim\sigma_z^2/\Sigma$, where $\Sigma$ is the local surface density.  
Inserting this into our scaling for $\tau$ gives
$$\tau\propto{{\sigma_z^2}\over{\cos^2 i}}.$$ This, of course, is 
Gould's (1995) analytic result.  Gould's point is that for
disk-disk lensing, to lowest order, the distance of the LMC and the
total mass (rotation curve) are irrelevant; that is, the only directly
observed quantities that seem to matter are the velocity dispersion and
axis ratio.  It is important to keep in mind that this
conclusion is predicated upon the validity of the self-gravitating,
thin-disk, steady-state solutions to the Jeans equations, which
Weinberg (1999) has argued may not be applicable to the LMC.  
We feel that it is better to base microlensing estimates 
upon parameters like the scale height, that are directly tied to the 
spatial density distribution, rather than quantities like the velocity
dispersion, which require questionable assumptions.

\begin{figure*}
\epsfysize=10.0cm
\centerline{
\rotate[r]{\epsfbox{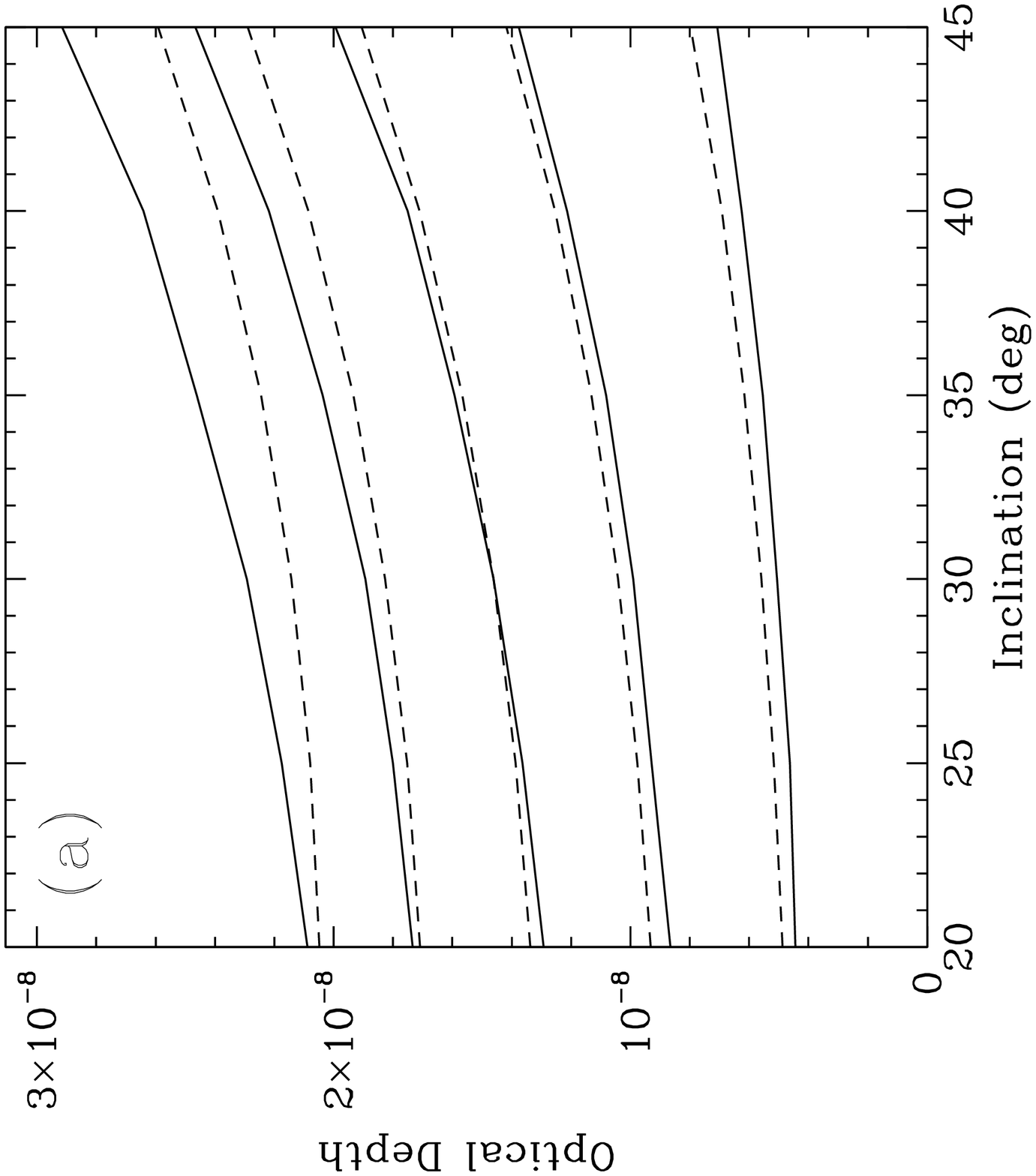}}
\rotate[r]{\epsfbox{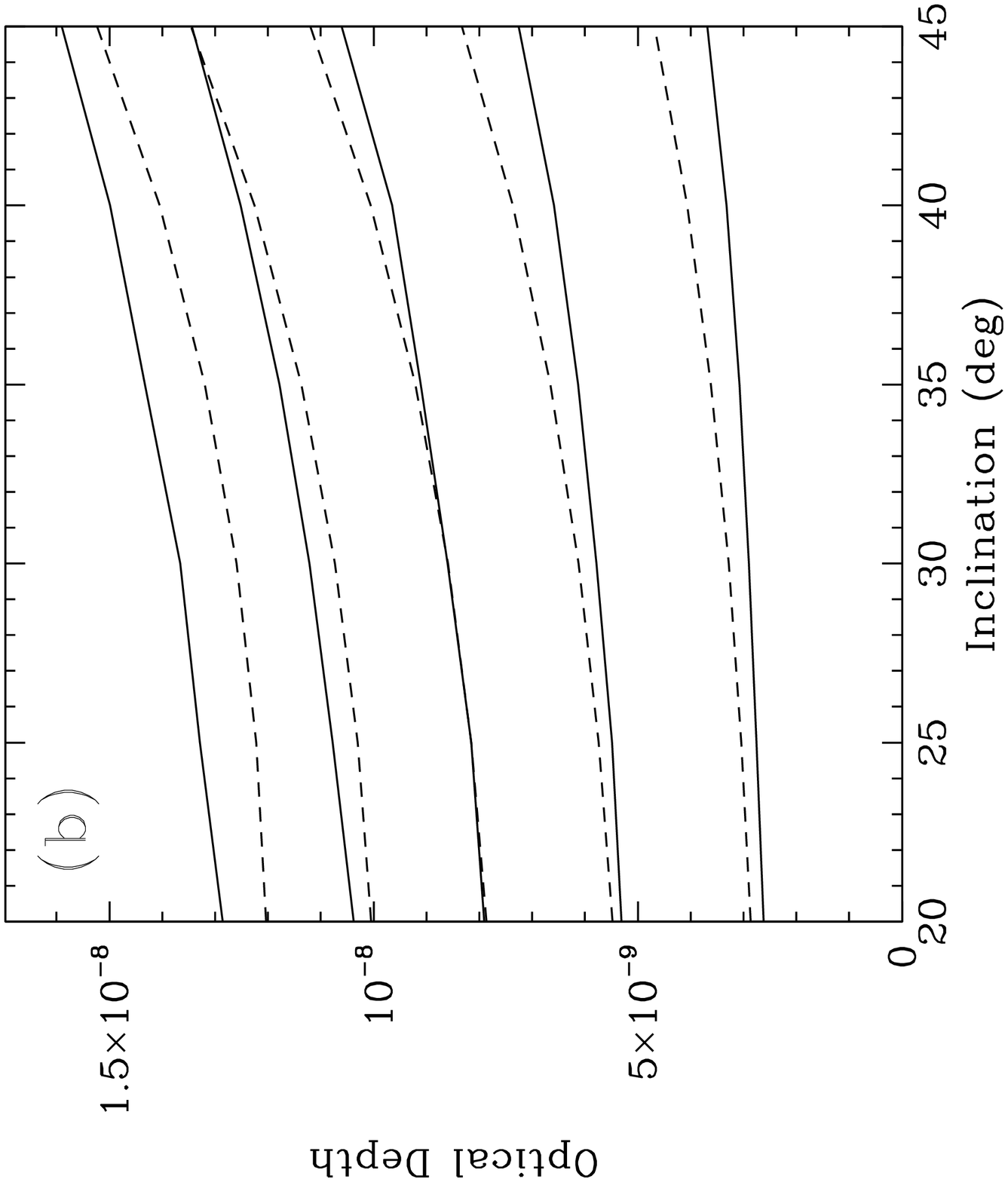}}}
\epsfysize=10.0cm
\centerline{
\rotate[r]{\epsfbox{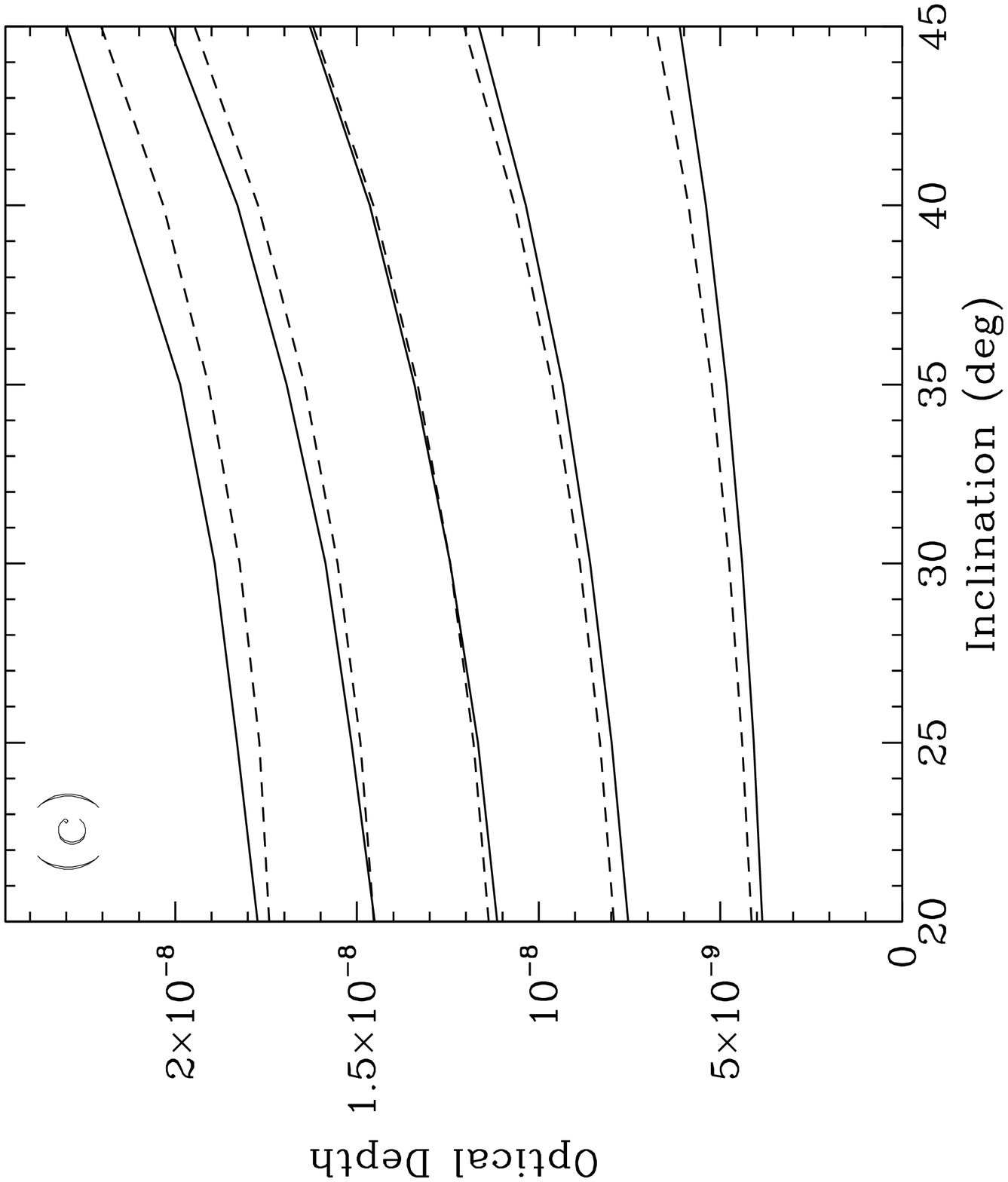}}
\rotate[r]{\epsfbox{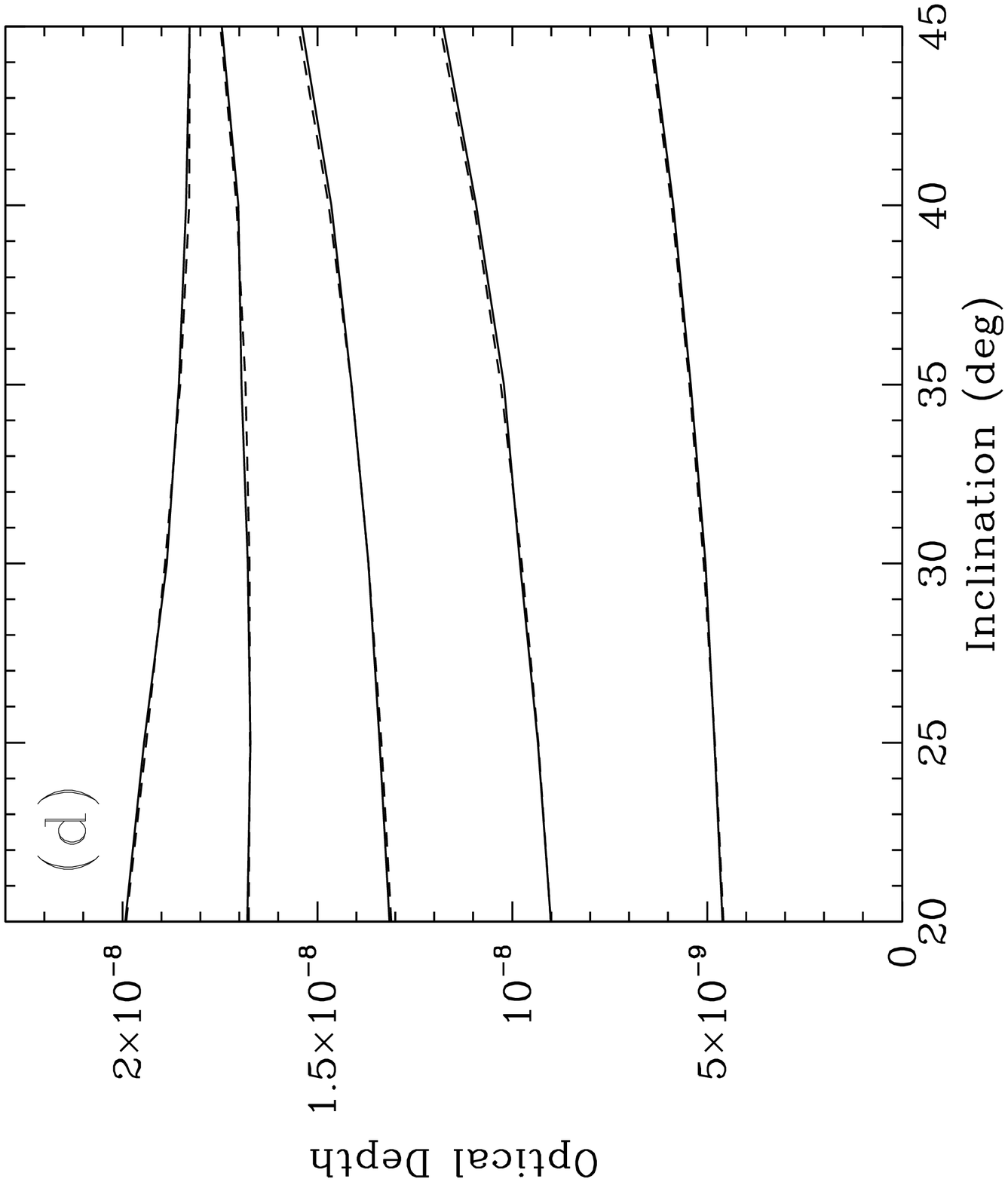}}}
\caption{
Variation of optical depth with model parameters.  Panel (a) is for
disk-disk, (b) is disk-bar, (c) bar-disk, and (d) bar-bar.  The solid
lines are the results of our numerical calculation of $\tau$, for
vertical scale height $z=0.1,0.2,0.3,0.4,0.5$ from bottom to top.  
All other parameters were set to their preferred values. The
dashed lines are the predicted values using the scalings described in
the text, for the same parameters. 
As expected, the scaling is most accurate for bar-bar, where both the
source and lens distributions are compact.
\label{figscalings}
}
\end{figure*}

The optical depths and expected scalings for the other three cases -- 
bar-disk, bar-bar, disk-bar -- can be computed with similar ease.  For
our preferred set of parameters, we find $\tau_{bd}=1.25\ten{-8},
\tau_{bb}=1.37\ten{-8}, \tau_{db}=8.7\ten{-9}$ (22 fields).  Again, we 
may also be interested in these optical depths for different parameter
values, so let's consider the scaling behavior of these integrals.  Now, 
we previously derived an approximate form by considering the limit of a 
compact source distribution and diffuse lens distribution.  What if we
reverse the situation, and instead imagine a compact lens distribution 
and extended source distribution?  For definiteness, consider a single lens
at the LMC center, lensing background stars with density profile
$\rho_s$.  In addition, let $\rho_s$ be strongly peaked about the LMC center.
Then the optical depth takes the approximate form 
$\tau\propto\int dl\rho_s(l)Ll/(L+l)\sim L^2\int\rho_sxdx$, the exact
same form we derived earlier, but with $\rho_s$ replacing $\rho_l$.
This should not be surprising, since in the limit $D_{OS}\approx D_{OL}$,
the Einstein radius becomes a function only of $D_{LS}$.  Since only the
relative distance from source to lens matters, $\tau$ becomes (in some sense)
symmetric in $\rho_s$ and $\rho_l$.  To sum up, when both the 
source and lens distributions are compact, but the source distribution is
more compact, we expect $\tau\propto L^2\int\rho_lxdx$, and when the
lens distribution is more compact then $\tau\propto L^2\int\rho_sxdx$.
For true self-lensing (disk-disk or bar-bar) these expressions are identical.

With these ideas in mind, we can now work out the approximate scalings.
Let's first consider bar-bar lensing.  The dimensionless integral 
in this case behaves, to leading order, like $\int e^{-Bx^2/2}xdx\sim B^{-1}$, 
where $$B=L^2\left[{{\cos^2 i}\over{z_b^2}}+\sin^2i\left({{\sin^2\psi}\over
{x_b^2}}+{{\cos^2\psi}\over{y_b^2}}\right)\right],$$
and $\psi$ is related to the bar's position angle on the sky by 
$\tan({\rm PA}_{bar}-{\rm PA}_{disk})=\tan(\psi)\cos(i)$.
Thus, $\tau$ scales roughly like 
$$\tau\propto{{M_{bar}L^2}\over{x_by_bz_bB}}.$$
Now we turn to the cross terms (disk-bar and bar-disk).  Clearly, the
bar Gaussian distribution is more compact than the disk double exponential
distribution, so we expect both of these terms to have the disk scaling.

Figure \ref{figscalings} shows the calculated optical depths 
averaged over 22 fields, as a
function of various parameters, as well as the predictions from the
scaling laws (normalized to match at the preferred parameters). 
In general the scalings are reasonably accurate.
The scalings work about as well for the 30 field set and the 82 field set.
For an order of magnitude estimate, one can use pure disk-disk with 
the sub-zeroth order scaling given above, namely
\begin{eqnarray}
\tau \approx 1.7 \cdot 10^{-8} &\left[{z_d\over{0.3 kpc}}\right] 
\left[{M\over{3\cdot 10^9 M_\odot}}\right] \times \nonumber \\
&\left[{{R_d}\over{1.6 kpc}}\right]^{-2}
\left[{{\cos(i)}\over{\cos(30^\circ)}}\right]^{-2}.\nonumber
\end{eqnarray}
For the 82 field sample the scaling is the same but the prefactor
is 1.05, while for the 30 field sample it is 1.33.  A complete average over
the LMC disk out to very large radii would give a prefactor of 0.72.

The total optical
depth for a particular set of model parameters is somewhat involved to
calculate. Since the spatial distribution of the various source populations is
different one cannot simply add together the mass weighted average optical
depths. Instead, one needs to add the optical depths weighted at the field
level and then calculate the total optical depth from the field values.
No combination of parameters within our ranges allows an optical depth
greater than $8.0 \cdot 10^{-8}$.  For the preferred values the optical
depth is $2.4 \cdot 10^{-8}$.  Note that this value is {\it ten times} 
smaller than the observed optical depth.

\begin{figure}
\epsfysize=10.0cm
\centerline{
\rotate[r]{\epsfbox{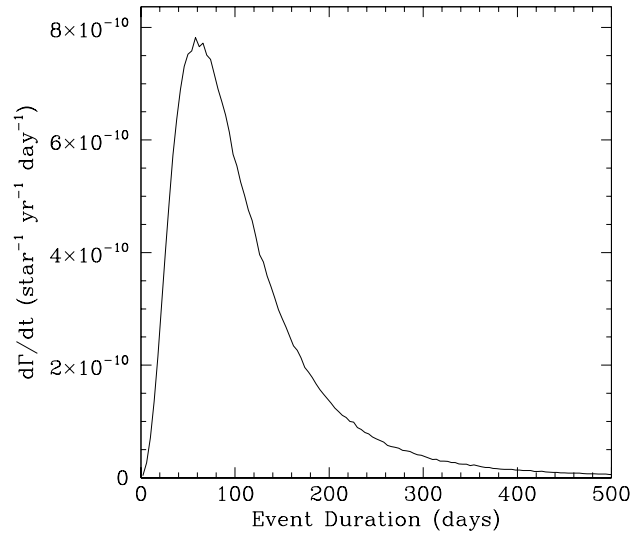}}}
\caption{
Timescale distribution averaged over 22 fields for our preferred model.  See
text for more details.
\label{figtimescale}
}
\end{figure}

\begin{figure}
\epsfysize=10.0cm
\centerline{
\rotate[r]{\epsfbox{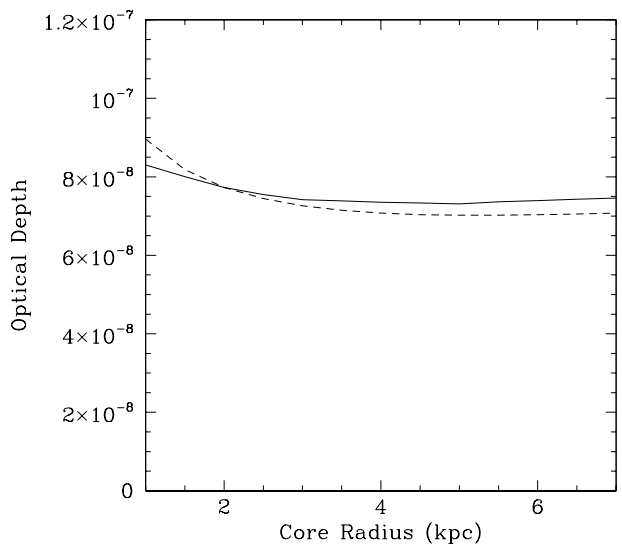}}}
\caption{
Optical depth as a function of halo core radius. The halo is taken to be
100\% MACHOs with a mass of $6.0\ten{9}\Msol$ within 8 kpc. The source
distribution is our preferred disk/bar model. Note how insensitive the
optical depth is to the core radius. The dashed line shows the scaling
described in the text.
\label{fighalotau}
}
\end{figure}

\subsection{Timescales}
Figure \ref{figtimescale} shows the timescale distribution for our
preferred disk/bar model, not including any LMC halo contribution,
using the mass function from Gould et al. (1997).
The efficiency weighted average event duration 
$\langle\hat{t}\rangle=101$\ days.  This is consistent with the observed
$\langle\hat{t}\rangle$ of 84 days \cite{LMC2}, given the observational
uncertainties and our lack of knowledge of the precise details of the
mass function and velocity distribution.  This profile is fairly uniform
over the face of the LMC.

\subsection{LMC Halo optical depth}

We have also calculated the optical depth due to a possible LMC MACHO
halo. The optical depth is shown in Figure \ref{fighalotau} as a function
of the halo core, $a_h$. If the mass of the halo is varied the optical
depth scales linearly. The values shown are for a $6\ten{9} \Msol$ LMC
halo 100\% composed of MACHOs.  We find values of the optical depth
between $\sim 7.5 \ten{-8}$ and $\sim 8.5 \ten{-8}$ depending on
parameters.  It is clear that a halo type configuration is much more
effective at producing optical depth than the disk/bar.  Now, there are at
least two possible types of LMC halos, both of which could, conceivably,
be present simultaneously.  First, a dark matter halo, common in dwarf
galaxies, is possible.  The composition of such a halo should be similar
to the composition of the Milky Way halo (i.e. unknown!).  If the MW halo
has a fraction $f_M$ of MACHOS (the remainder presumably consisting of
some exotic non-baryonic material), then the LMC halo might have a similar
fraction.  If so, then microlensing of the LMC halo lenses would
constitute discovery of dark matter, but the implied halo fraction would
depend upon the mass of LMC halo \cite{wein98,evans}.  
That is, the predicted optical depth of the MW halo
plus LMC halo would be roughly $\tau \simeq f_M (4.7\ten{-7} +
[0-2.3]\ten{-7})$, and so the effect of including a dark LMC halo would be 
to reduce the derived MACHO halo fraction by 
$\Delta f_M/f_M = -[0-2.3] / (4.7+[0-2.3]) = -[0\% - 33\%] $.
Using the current estimate $f_M$ of 50\% \cite{LMC2}, inclusion of an 
LMC halo lowers $f_M$ to somewhere in the range [33\% - 50\%].

The other type of possible LMC halo is a stellar halo with a luminosity
function similar to that in the disk.  This could consist of stars
stripped from the disk (Zaritsky \& Lin 1997; Weinberg 1999)
or something corresponding to the spheroid of the Milky Way. 
This is the halo of interest in creating a non-dark matter explanation
for LMC microlensing.  As discussed above, the mass of such a halo is 
tightly constrained by numerous observations.

As before, we can work out an approximate scaling for the optical depth of 
LMC halo lensing.  The halo distribution is definitely less compact than the 
source distribution, so $\tau\propto L^2\int\rho_lxdx$.  Let $a_h$ 
be the halo core radius, $r_t$ the tidal radius (actually truncation radius), 
$\rho_0$ the halo central density, and $L$ the distance to the LMC.  Then 
the optical depth should
scale like 
$$\tau_h\propto\rho_0a_h^2\left[{1\over 2}\log\left(1+{{r_t^2}\over{a_h^2}}
\right)+{{\tan^{-1}(r_t/a_h)}\over{L/a_h}}-{{r_t}\over L}\right].$$

\begin{table*}
\begin{tabular} {lccccccccl}
Source/Lens geometry & \multicolumn{3}{c}{Relative Weight}&~~~
&\multicolumn{3}{c}{Preferred Values} &~~~& Allowed Range\\
&22 & 30 & 82&& 22 & 30 & 82&&(22 fields)\\
\hline
disk/disk	& 0.61 & 0.67 & 0.79 && 1.46 & 1.34 & 1.04 && 0.23-5.81\\
disk/bar	& 0.61 & 0.67 & 0.79 && 0.87 & 0.72 & 0.39 && 0.11-4.07\\
bar/disk	& 0.39 & 0.33 & 0.21 && 1.25 & 1.24 & 1.23 && 0.40-4.13\\
bar/bar		& 0.39 & 0.33 & 0.21 && 1.37 & 1.36 & 1.33 && 0.32-4.00\\
total bar+disk  & 1    & 1    & 1    && 2.44 & 2.24 & 1.67 && 0.47-7.84\\
(disk+bar)/dark halo & $f_M$ & $f_M$ & $f_M$ && 7.75 & 7.73 & 7.18& & 0 - 22.6 \\
\hline
\end{tabular}
\caption{Optical depths for LMC self lensing in units of $10^{-8}$, averaged
over the 22 fields of Alcock et al. (1997a).  The dark halo result is for 
a MACHO fraction $f_M=1$.  Note that the relative weights apply only to the
the preferred set of parameters.
}
\end{table*}

\subsection{Total Optical Depth}
All of the above components combine to give the total predicted optical
depth for LMC/LMC self lensing.  This averaging is not completely trivial
since the density of source stars is different in each population.  To
find an average optical depth for a set of model parameters and a set of
observed fields, the optical depths for each population should be
multiplied by the source density at each field location and then averaged
with this weighting.  Even more realistically, observational effects such
as stellar crowding and observation strategy will cause the monitored
source objects to differ from the underlying stellar sources, and the
detection efficiency of each field to vary, so additional corrections for
each field should also be made.  We blissfully ignore all such
observational effects, and assume that our model source distribution
approximates the true observed source distribution with uniform source
exposure and detection efficiency.  Table~4 shows a summary of the optical
depths for the various populations discussed above
and also the averaged totals.  The ranges of parameters shown in Table~3
were used.  We see that for our preferred parameters a total optical depth
due to known LMC stellar populations
of $2.44\ten{-8}$ is found, with values between 0.47 and 7.84$\ten{-8}$ 
lying in our acceptable range.

\subsection{Variation of Optical Depth across the face of the LMC}
One potentially powerful way of distinguishing Milky Way (MW) halo 
microlensing from
LMC self-lensing is to compare the spatial distribution of the observed
microlensing events with the predictions of LMC and halo models
\cite{LMC2}.  
For microlensing events due to a MW halo population of lenses, the lens
population is uniform across the source distribution, so one expects the 
events to be
distributed in proportion to the LMC source density times the experimental
efficiency.  For LMC/LMC self lensing, both the sources and lenses are
distributed like the LMC stars, so there should be a more rapid drop off
of measured optical depth at large distances from the LMC bar.

\begin{figure*}
\epsfysize=25.0cm
\centerline{
\rotate[r]{\epsfbox{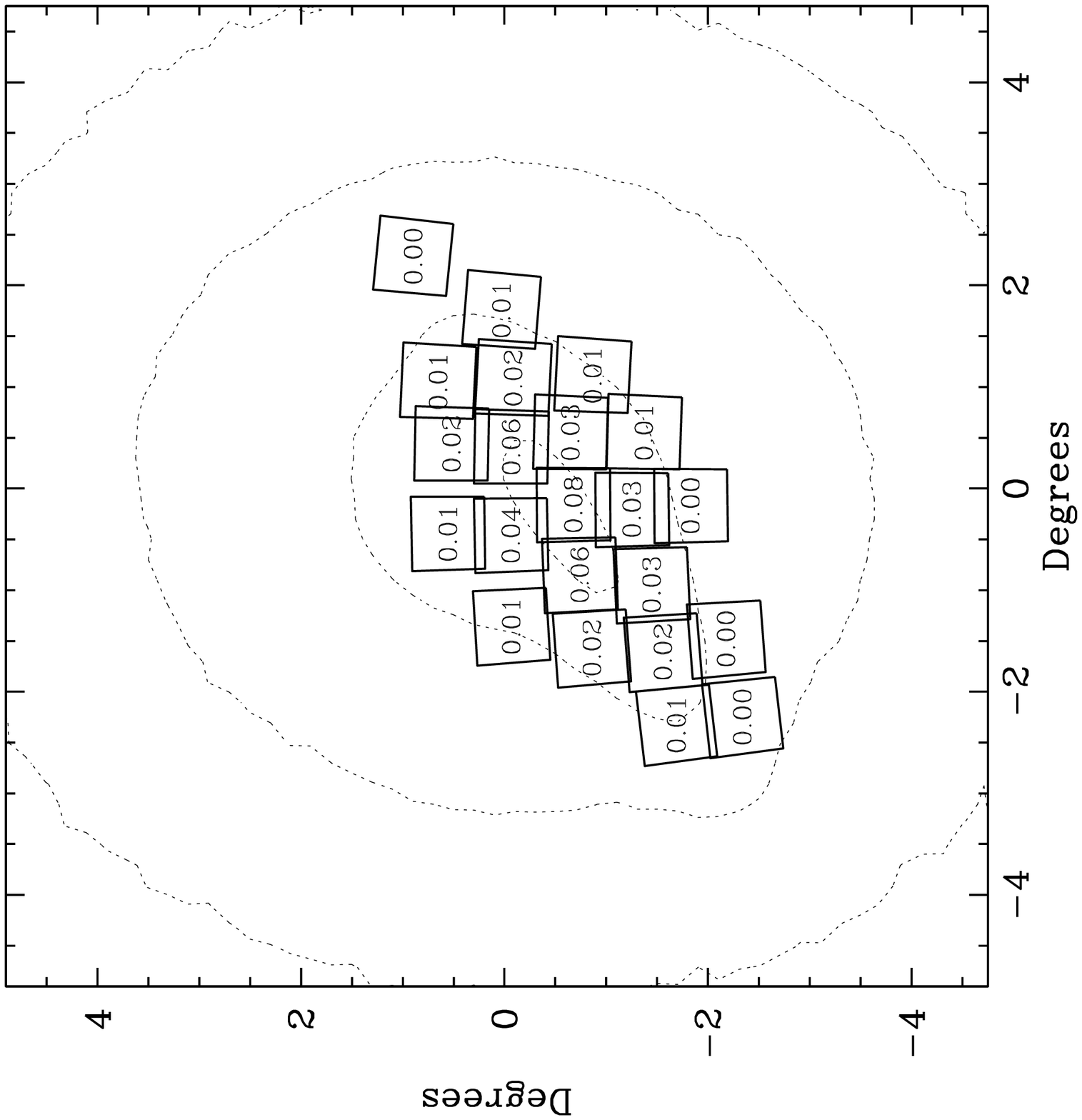}}}
\caption{
The dotted contours depict the optical depth times source number density as 
a function of position 
on the sky for LMC self-lensing, for our preferred model with both disk and
bar.  The contours are spaced by decades.
The 22 fields are overlaid, along with the expected number of events using our
preferred model, an exposure of $1.82\ten{7}$ star-years, and the detection
efficiencies of Alcock et al. (1997a).  We find a total of 0.44 expected 
events.
\label{figvariationdisk}
}
\end{figure*}

\begin{figure}
\epsfysize=10.0cm
\centerline{
\rotate[r]{\epsfbox{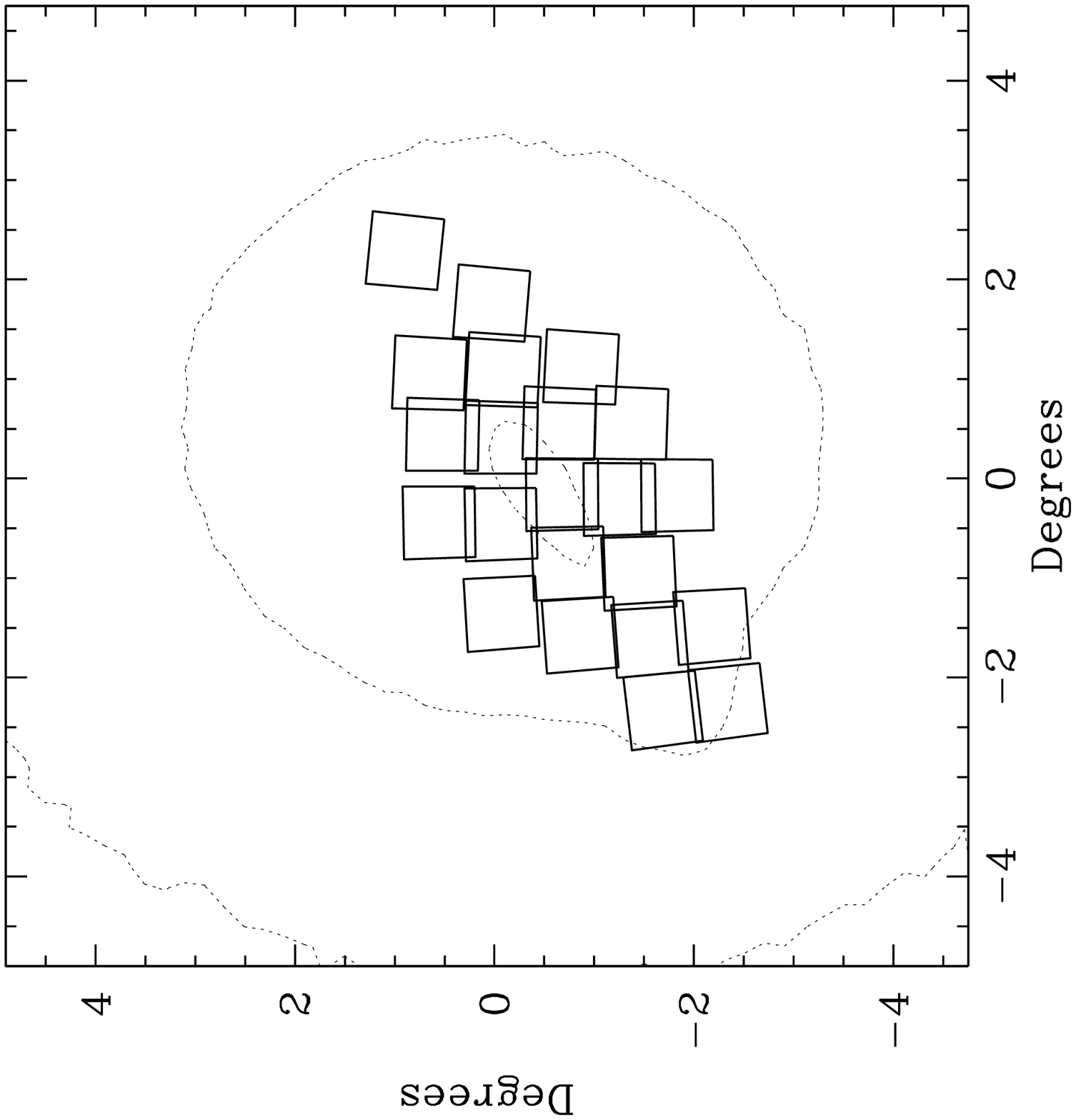}}}
\caption{
Same as figure \ref{figvariationdisk}, but for lensing by the LMC halo.
Note that the distribution is much less compact than the corresponding 
disk/bar result; that is, the contours are much more widely spaced out.
\label{figvariationhalo}
}
\end{figure}

In Figure \ref{figvariationdisk} we show the predicted distribution of
disk/bar optical depth (times the source density in the model) as a
function of RA and declination across the face of the LMC for our
self-lensing model, employing the preferred parameters. It is clear that
the optical depth drops off rapidly with radius.  We note that even a few
events in the regions far from the bar can rule out the LMC/LMC self
lensing hypothesis, if the LMC lens population is accurately modeled as a
disk/bar. Figure \ref{figvariationhalo} shows the same for LMC halo lensing.
Note that the halo optical depth is much less concentrated than the
corresponding disk/bar result.  Interestingly, there is
a slight east-west asymmetry for LMC halo microlensing due to the
inclination of the disk.  Although such an asymmetry would be virtually
impossible to detect experimentally, in principle it could be used to
discriminate between LMC halo and MW halo microlensing.

Looking at Figures \ref{figvariationdisk} and \ref{figvariationhalo}, we
see that the disk/bar distribution is qualitatively distinct from the halo
distribution.  We can quantify this observation using a simple measure.
We write $\tilde{\theta}\equiv\langle\theta_{ij}
\rangle$, where $\theta_{ij}$ is the angle on the sky between the 
location of events $i$ and $j$, and the average is over all pairs.
$\tilde{\theta}$ is a statistic that measures the average separation
of events, and therefore the extent of the spatial distribution of
events.  It is easy to compute the experimental value,
$\tilde{\theta}_{obs}$, from the observed events.  It is similarly
easy to compute the values predicted by any given model.  Since the
main feature of the event distribution that should help rule out
models is the central compactness, we expect that $\tilde{\theta}$
should measure whether a model can reproduce the observed event
distribution, in the same sense that a KS test would.  Unfortunately,
the paucity of actual events may limit our ability to rule out models
based upon the observed event distribution.  In addition, to be useful
the monitored sources must span a sufficiently wide area, since
obviously we will be unable to discern any intrinsic central
concentration if the data sample only a small swath of the sky.

\begin{figure}
\epsfysize=10.0cm
\centerline{
\rotate[r]{\epsfbox{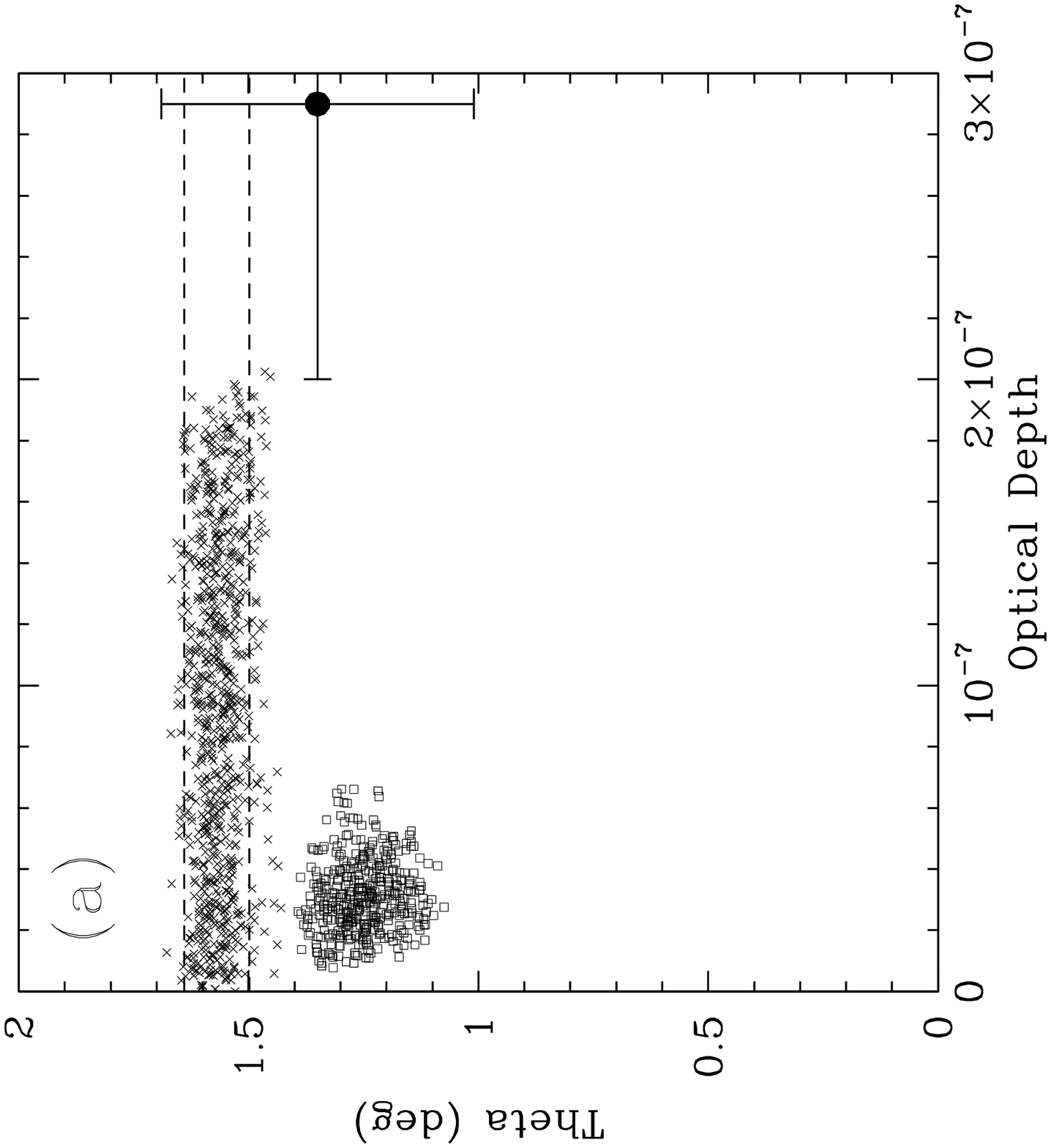}}}
\epsfysize=10.0cm
\centerline{
\rotate[r]{\epsfbox{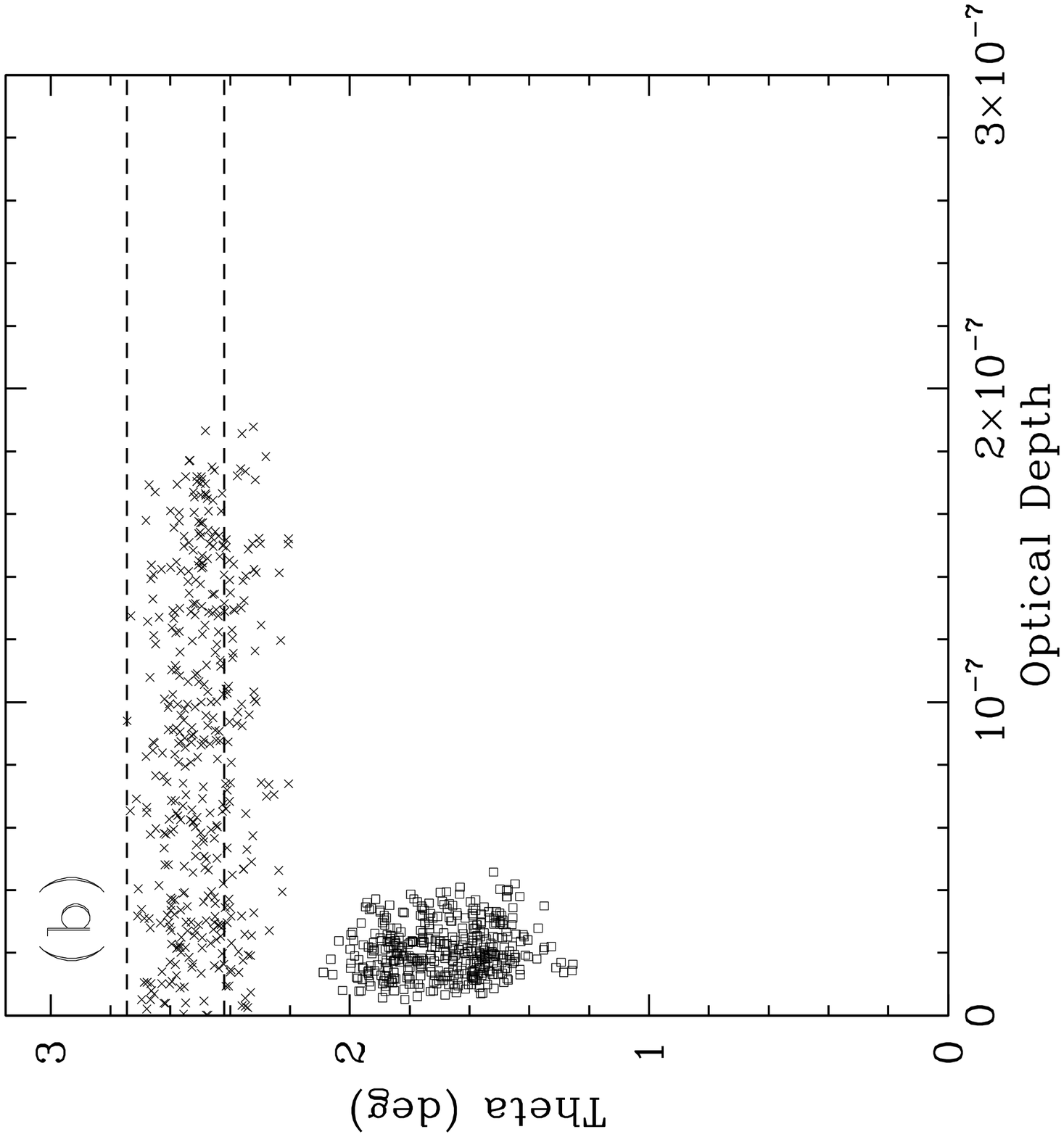}}}
\caption{
$\tau$ {\it vs.} $\tilde{\theta}$.  Panel (a) shows the range of
$\tau$ and $\tilde{\theta}$ for the 22 field set, and (b) shows the same
for the 82 field set.  The open boxes are for the disk/bar, and the
x's are for the LMC halo.  The points were randomly selected uniformly
in parameter space, within the allowed ranges.  The dashed lines show
the range for MW halo lensing.  The point with error bars corresponds
to the MACHO year 2 events (Alcock et al. 1997a).  Note the increased 
separation of the disk/bar and halo distributions for the 82 field set.
\label{figthetat}
}
\end{figure}

We have computed $\tilde{\theta}$ for our models using both the 22
field sample and the 82 field sample.  In order to give an estimate of
the allowed range of $\tilde{\theta}$, we have plotted it for numerous
random parameter sets picked uniformly in parameter space.  Figure
\ref{figthetat} shows a scatterplot of $\tau$ versus $\tilde{\theta}$
for LMC disk, LMC halo, and MW halo models.  For the 22 field sample
depicted in panel (a) the LMC disk/bar (x's) and the LMC halo
(circles) models are resolved in both $\tilde{\theta}$ and $\tau$,
though the experimental uncertainties on
$(\tilde{\theta}_{obs},\tau_{obs})$ plotted for the MACHO LMC 2 year
data set do not allow them to be distinguished.  We see that the
selection effect of small sky coverage, in conjunction with low-number
statistics, does not yet allow a clear choice of model based upon the
event distribution.  On the other hand, the 82 field sample plotted in
panel (b) demonstrates strong separation of the model classes in both
$\tilde{\theta}$ and $\tau$.  The 30 fields result is quite similar to
the 22 fields plot, with a maximum range in $\tilde{\theta}$ of
$\simlt 2^\circ$ for halo lensing.  Clearly, future data with
increased sky coverage and more events will be a strong discriminant
between disk models and halo models.  Note, however, that the plots
also indicate that the LMC halo and MW halo models will probably
remain degenerate.

We now turn to a discussion of
other self-lensing models that have been proposed recently.

\subsection{Other Models}
\begin{table*}
\begin{tabular} {lccccccl}
Paper & $M_{d+b}$ & $M_{stellar~halo}$& $z_d$ & $\cos(i)$ & $\tau_{paper}$ 
&$ \tau_{us}$ \\ 
\hline
Sahu 1994&$2\ten{9}M_\odot$&0&$\sim 0.2$kpc&0&$5\ten{-8}$&$5.3\ten{-8}$ \\
Gould 1995&$\sim 1.2\ten{9}M_\odot$&0&$\sim 0.2$kpc&$27^\circ$&
$\simlt 10^{-8}$&$\sim 7.6\ten{-9}$\\
Alcock, \etal 1997a &$6.8\ten{9}M_\odot$&0&0.25 kpc&$30^\circ$&$3.2\ten{-8}$&
$3.2\ten{-8}$\\
Aubourg 1999 &insignificant&$\sim 1.4\ten{10}M_\odot$&-&-&$1.3\ten{-7}$&
$1.2\ten{-7}$\\
Weinberg 1999 &$10^{10}M_\odot$&0&$\sim 0.4$kpc?&$45^\circ$&$1.4\ten{-7}$&
$1.4\ten{-7}$& \\
This work &$3\ten{9}M_\odot$&0&0.3 kpc&$30^\circ$&$2.44\ten{-8}$&
$2.44\ten{-8}$& \\
\hline
\end{tabular}
\caption{Summary of LMC self lensing optical depths results for various groups.
See text for more explanation.  Gould's, Aubourg's \& Weinberg's
results are all for a single line-of-sight, which overestimates the 22
field averaged optical depth by $\sim 50\%$.  Sahu's model was pure
bar, with no disk.  Gould's optical depth was expressed in terms of
the vertical velocity dispersion; we chose values of the disk mass and
scale height which roughly give that dispersion.  It is unclear what
scale height corresponds to Weinberg's calculation, but 0.4 kpc is
a reasonable estimate.}
\end{table*}

The models of Sahu (1994a,b), Gould (1995),
Alcock, \etal (1997a), Aubourg, \etal (1999), are all contained
within the framework we have explored.  The main differences between
the results of these different workers come from different choices of
LMC model and LMC model parameter.  In Table~5 we show a summary of
the LMC self-lensing results of several previous workers along with
the parameters they chose.  We also show an approximation of their
models within our framework.  Note in every case, the predicted
optical depth can be found using our formulas and models, and their
LMC parameters.  This is even true for the sophisticated N-body
calculation of Weinberg (1999); substituting in his final values of
parameters gives nearly the same answer as he found from his
interacting and tidally disrupted model.  We conclude that
disagreements about the values of optical depth can be traced to
disagreements about parameter choices.  Workers with values of optical
depth above $10^{-7}$ all chose parameters outside of
our allowed range. 

Clearly, to settle these questions, better observations of the stellar
components of the LMC and its environs are needed.  Direct evidence for
a stellar component with
an extended or halo geometry would be a key to confirming a
non-dark matter explanation for LMC microlensing.  We now discuss some
of the individual models in more detail.

Aubourg \etal~(1999) have suggested an LMC model which would produce a
self lensing optical depth of $\sim 1.3 \cdot 10^{-7}$. If true, this
would appear to solve the LMC microlensing problem. Unfortunately,
there are serious objections to be raised against the stellar lensing
population they employ, since there are no known tracer populations.
In particular, the lenses in their model, which are ordinary stars,
typically have masses between 0.1-1 $M_\odot$, and the vast majority
are arranged in a spherical (axis ratio $\sim$ 0.9) isothermal
distribution with velocity dispersion $\sim 45$ km/s.  This profile
and velocity are inconsistent with a multitude of tracers of the old
population.  Among these are the OLPV results of Bessel et al. (1986)
and Hughes et al. (1991), the metal poor giants of Olszewski et
al. (1993) and halo-type CH stars of Cowley \& Hartwick (1991).  While
arguments could be made that any individual tracer is not really old
or does not represent the old population as a whole, taken together
the evidence against the bulk of the mass of the LMC being in the form
of an old {\it stellar} halo is strong.  Indeed, the distribution of
old clusters \cite{schommer} and the RR Lyrae star counts 
\cite{MACHO9million} are particularly telling as they are almost
certainly an ancient population. The Aubourg et al. model thus appears
to be at odds with current observations of the LMC disk.
Additionally, it is unclear that the process they invoke for
populating their spheroidal component, stochastic heating of the disk
by inhomogeneities in the disk itself, would be capable of ejecting
upwards of 80\% of the LMC disk mass into a far less centrally
concentrated pseudosphere.

We can recast their model, however, in a potentially more palatable
form.  The MACHOs conjectured to reside in the halo of {\em our}
Galaxy have mass of about 0.2-0.8 M$_\odot$, and are modeled in a
spherical pseudoisothermal distribution.  Note the striking
resemblance between the conjectured Milky Way MACHOs and the LMC
lenses proposed by Aubourg et al.  Whether one chooses to call these
undetected objects stars or MACHOs becomes a matter of semantics; we
see that the Aubourg et al. model is identical in practice to a MACHO
halo around the LMC.  As expected, therefore, their results match our
calculation for disk-halo lensing.

Very recently, Weinberg (1999b) has suggested that a substantial
portion (perhaps all) of the LMC microlensing might be due to LMC disk
self lensing. He models the LMC self-consistently with a sophisticated
N-body code and finds that the effect of the time-varying Galactic
tidal forces is to puff up the LMC disk by a factor ($\simgt 2\times$) 
without noticeably increasing the velocity dispersion.  This is important 
since the optical depth depends strongly on scale height.
Indeed, if this simulation does in fact resemble the LMC's history,
then it argues against reliance upon thin-disk equilibrium
solutions to the Jeans equations, {\it a la} Gould (1995).  Weinberg's
reported optical depth is $1.4 \cdot 10^{-7}$, which compares
favorably with the recent estimates from the MACHO collaboration
($\sim 2.0 \cdot 10^{-7}$) \cite{will96}.  A few points need to be
made regarding this result.  First, this optical depth appears to be
calculated for the line of sight to the center of the LMC, instead of
averaging over the observational fields.  This will yield results
biased upward by a factor of about 1.5. Second, Weinberg assumes an
inclination of 45$^\circ$.  Using the more likely preferred value of
30$^\circ$, yields a further 15\% reduction (see Weinberg's figure
12).  Finally, the disk mass taken in his study, $10^{10}\Msol$ appears 
unrealistically high.  As discussed above, (and matching nicely with 
Weinberg's own calculations in his Appendix A) the {\em disk} mass is 
unlikely to be above $5 \cdot 10^{9}\Msol$.  Taking all of these 
adjustments into account the optical depth is reduced to 
$\sim 4\cdot 10^{-8}$, falling in line with the values calculated in this 
work.

Finally, we should note that Zhao (1998) has suggested that an intervening
dwarf galaxy similar to the Sagittarius dwarf could be responsible for the
LMC microlensing events.  Searches for the RR Lyrae stars that should be
contained in such a dwarf turned up negatively \cite{dantedwarf}, but
Zaritsky \& Lin (1997) evaded these search limits by hypothesizing the
existence of either a dwarf galaxy very near the LMC itself, or perhaps a
tidal tail pulled from the LMC by a close encounter with the SMC.  They
claimed detection of a population of stars from this intervening entity.
This result has been disputed in several ways by several groups.  Gould
(1998) and Bennett (1998) claim that the optical depth due to the Zaritsky
\& Lin population is insufficient to explain the microlensing results.
Beaulieu \& Sackett (1998) claim that the Zaritsky \& Lin stars are
ordinary LMC stars that are brighter due to stellar evolution, and thus do
not represent an intervening population.  The discussion continues
\cite{newzaritsky,newgoulddwarf}, and at this point, while the question has 
not been definitively settled, the case for an intervening dwarf looks rather 
weak.

\section{Discussion}

We have seen that pure LMC disk/bar self-lensing models appear
incapable of producing the measured optical depth of
$\tau\approx2.9\ten{-7}$.  Given the current state of observations of the
LMC, the most likely self-lensing optical depth is an order of
magnitude too small to account for the observed events.  A reasonable
range of self lensing optical depths is 0.47 - 7.84$\ten{-8}$ 
depending upon model parameters.  
We have shown that halo models can reproduce the optical
depth, {\em if} we are allowed to push the model's parameters to their
extremes.  We pointed out that numerous observations already limit our
ability to push the parameters very far.  
In order to invoke a self-lensing explanation of LMC microlensing,
observation of a sufficient number of stars exhibiting the characteristics
of an extended halo seems crucial.
Better observations of the LMC that more strongly constrain the disk
scale height, inclination, disk mass, total mass,
and velocity dispersion also are important
as they will reduce the allowed range of optical depth.
Especially important are measurements of velocity dispersions and spatial
extent of old populations such as RR Lyrae.

We then discussed how
consideration of the distribution of optical depth over the face of
the LMC can help further distinguish between models.  In particular,
LMC disk/bar lensing will produce events clustered around the LMC
center, while LMC halo lensing will produce a more diffuse
distribution of events.  We introduced a new
clustering statistic, and showed that more microlensing observations 
distributed over the face of the LMC can be very useful in identifying
LMC disk/bar self lensing.  However, distinguishing between an LMC halo
and the Milky Way halo will probably not be possible using this method.

Finally, we note that if the distance of some of the lenses could be directly
determined, the puzzle could be solved.  There are several ways to
do this using microlensing fine-structure, and the distance to at least one
binary SMC lens has been well determined \cite{machosmc,erossmc,planetsmc}.  
One LMC binary lens (LMC-9) for which such a determination was possible 
was unfortunately not sampled well enough to allow
a secure distance determination \cite{macho-binary}, but
continued monitoring should eventually allow some distances to be found.
Perhaps the most secure distance determination would come from astrometric 
parallax effects in microlensing.  NASA's Space Interferometry Mission (SIM), 
scheduled for launch about 2005, should have the astrometric resolution to
make definitive parallax measurements \cite{pac98,boden98}.

It has been argued that the SMC events along with LMC-9 strongly
support the notion that all of the events are due to self-lensing. In
this view, the next LMC binary event should definitively decide
between halo and self lensing scenarios.  This may not be the case.
First, the SMC is {\em expected} to have a high self-lensing rate
regardless of the nature of the LMC lenses.  Coupled with the
uncertain interpretation of LMC-9, the case for LMC self lensing is
poorly supported at present.  In this light, the next LMC binary event
alone will probably not suffice to locate the bulk of the lenses, even
if it is found to be an LMC lens.  First, we've seen that LMC self
lensing may well contribute of order 10-20\% of the lensing, so that
some self-lensing events are expected.  Secondly, there may well be
fewer binaries in the Halo than the LMC, inducing a possible selection
effect in favor of LMC lenses.  It will probably require multiple
future distance determinations to settle this matter.

Given the importance of the LMC microlensing interpretation to the dark
matter question, many of the potential observational efforts described
above are being attempted.  We are hopeful that the interpretation
of LMC microlensing will become clear within the next few years.

\section*{Acknowledgements}
It is a pleasure to thank Thor Vandehei for numerous enlightening
and stimulating discussions on this and related matters.  
We acknowledge support from the U.S. Department of Energy, under grant
DEFG0390ER 40546, and from Research Corporation under a Cottrell Scholar 
award.

\end{document}